\documentclass[a4paper,oneside,11pt]{article}
\usepackage{epsfig,multicol}
\usepackage{graphicx}
\usepackage{amsmath,amssymb}
\usepackage{mathrsfs}
\usepackage{graphics,xcolor}
\usepackage{caption}
\usepackage{subcaption}

\numberwithin{equation}{section}

\begin{document}
\setcounter{page}{1}

\begin{center}
  {\Large \bf Marchenko method with incomplete data and singular nucleon
              scattering}
\end{center}

\vskip 19.6ex

\begin{center}
Mahmut Elbistan$^1$\footnote{\tt elbistan@impcas.ac.cn}\,,\ \  
Pengming Zhang$^1$\footnote{\tt zhpm@impcas.ac.cn} 
\ \ and\ \ 
J\'anos Balog$^{1,2}$\footnote{\tt balog.janos@wigner.mta.hu}
\\
\vskip 3ex
$^1$ {\it Institute of Modern Physics, 
Chinese Academy of Sciences,\\ Lanzhou 730000, China}\\
\vskip 1ex
$^2${\it MTA Lend\"{u}let Holographic QFT Group, Wigner Research Centre} \\
{\it H-1525 Budapest 114, P.O.B. 49, Hungary}\\

\end{center}


\vskip 14.5ex

\noindent
We apply the Marchenko method of quantum inverse scattering to study
nucleon scattering problems.
Assuming a $\beta/r^2$ type repulsive core
and comparing our results to the Reid93 phenomenological potential we
estimate the constant $\beta$, determining the singularity strength,
in various spin/isospin channels.
Instead of using Bargmann
type S-matrices which allows only integer singularity strength, here we
consider an analytical approach based on the incomplete data method, which is
suitable for fractional singularity strengths as well. 

\newpage


\newcommand{\beq}{\begin{eqnarray}}
\newcommand{\eeq}{\end{eqnarray}}
\newcommand{\ee}{\end{equation}}



\section{Introduction and motivation}
\label{motiv}

\begin{figure}
\begin{flushleft}
\vskip -30mm
\leavevmode
\centerline{\includegraphics[width=7cm]{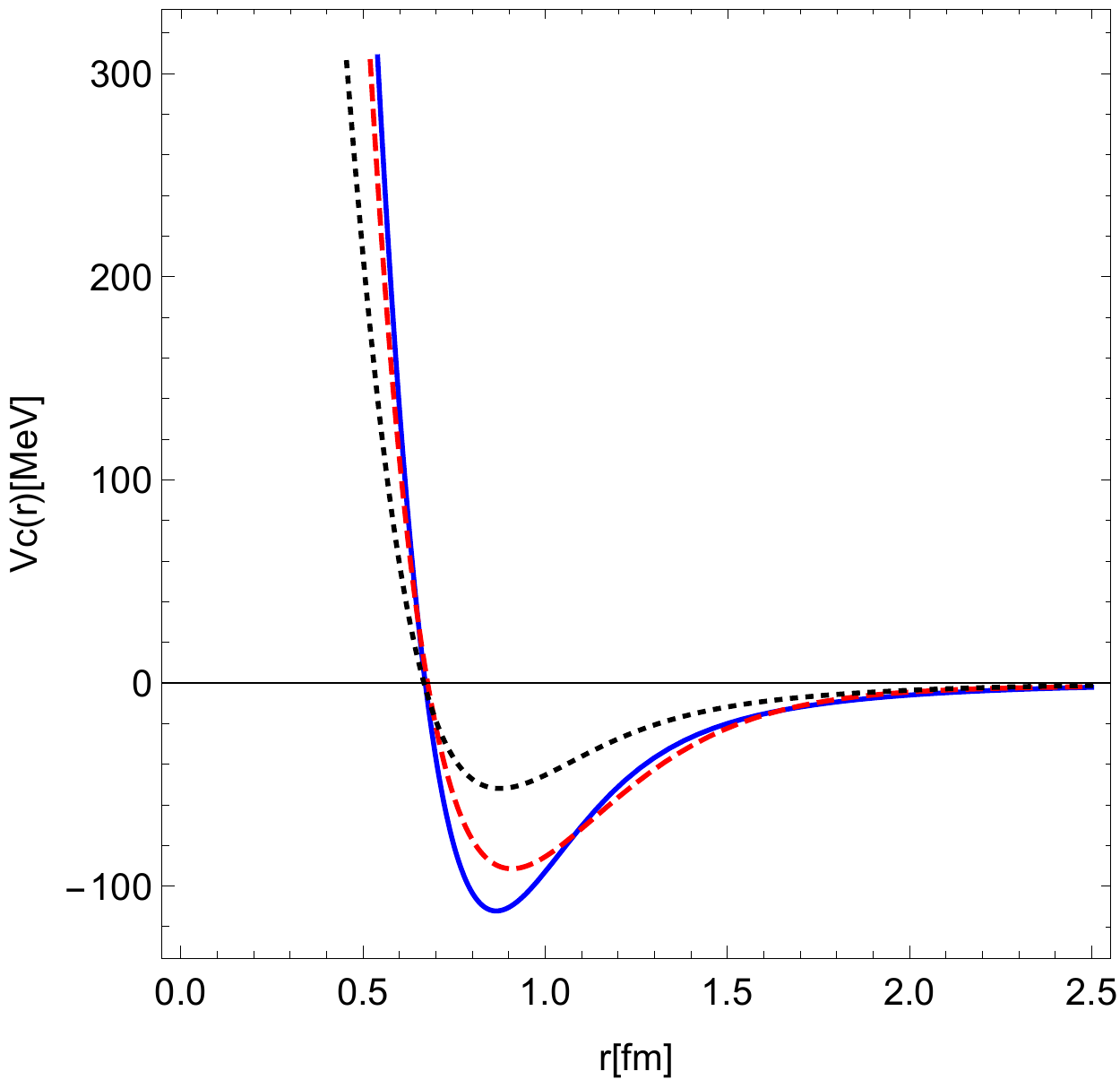} }
\end{flushleft}
\vskip -0.5cm
\caption{{\footnotesize
The phenomenological nucleon potential in the $^1S_0$ channel. The
solid (blue) line is the AV18 fit \cite{NuPo3}, the dashed (red)
line is the Reid93 fit \cite{NuPo2} and the dotted line (black)
is the CDBonn potential \cite{NuPo1}.
}}
\label{pheno}
\end{figure}

The characteristic features of the phenomenological nucleon
potentials \cite{NuPo2,NuPo3,NuPo1}, shown in Fig. \ref{pheno},
are well known. The force at medium to long range is attractive; this feature
is due to pion and other heavier meson exchange. The strong repulsive core
of the potential at short distances had no satisfactory theoretical
explanation until recent advances in lattice QCD simulations made possible
to determine the potential in fully dynamical lattice
QCD \cite{Ishii:2006ec,AokiRev}. The results of this first principles
calculation resemble the phenomenological potential, including its repulsive
core. The short distance behaviour of the potential was subsequently studied
also in perturbative QCD. The results of the perturbative
calculations \cite{PT1,PT2} show that at extremely short
distances the potential behaves as $1/r^2$ (up to log corrections characteristic
to perturbative QCD). Calculations in holographic QCD \cite{HoloQCD}
also give a similar inverse square potential at short distances.

Although the recent theory of low energy nuclear interactions is based on
effective chiral field theory (EFT) of mesons and nucleons (for a review,
see \cite{EFT}), the phenomenological potential remains important
as a source of intuition and is still often used in the study of multinucleon
systems and in the determination of the equation of state for dense
nuclear matter as starting point of quantitative work.

As can be seen in Fig. \ref{pheno}, the phenomenological potential is not
uniquely determined. Nevertheless, known versions more or less agree on its
main qualitative features at large and medium distances and substantially
deviate only at short distances (corresponding to higher energies).
From a purist viewpoint the notion of nuclear potential does not make
much sense much below 0.5 fermi for various reasons: the nonrelativistic
quantum-mechanical description based on the Schr\"odinger equation cannot
work beyond about 350 MeV laboratory (LAB) energy because it cannot
incorporate particle production; relativistic effects become important at
the corresponding energy range; finally the composite nature of nucleons
becomes relevant at distances comparable to their size. (By using EFT,
which is a relativistic quantum field theory, the first two difficulties 
are avoided but it cannot directly address the last problem either.) Therefore
a meaningful reconstruction of an effective nuclear potential
(more precisely a Hamilton operator in the Schr\"odinger equation) 
must be based on experimental data in the $0<E_{\rm LAB}<350\ {\rm MeV}$
energy range. This leads to the problem of quantum inverse scattering with
incomplete data.

We have investigated \cite{EZB} analogous problems in the $1+1$ dimensional
Sine-Gordon model. Since this relativistic model is integrable, there is
no particle production, but we can still ask the question whether an effective
potential exists which exactly reproduces the known scattering phase shifts.
We found that the answer is affirmative both in the centre of mass (COM) and
the LAB frame, but the price one has to pay is frame-dependence. However, we
also found that in this model the frame-dependence is weak, both at small and
large distances and the COM and LAB frame effective potentials are
qualitatively similar and numerically close also at medium distance. Thus an
approximate notion of effective potential makes sense, at least in this model.

In the theory of inverse scattering with incomplete data the lack of full
information on the scattering phase shifts is (partially) compensated by
other, additional pieces of information. In this paper we concentrate on the
singular core of the potential and assume it behaves for small $r$ as
\begin{equation}
U(r)\sim\frac{\nu(\nu+1)}{r^2}+{\rm O}(1)
\label{one}
\end{equation}
(in natural units), where the parameter $\nu$ is non-negative (repulsive core).
In a recent paper \cite{EZB2} we studied the singular behaviour of the nucleon
potential in the $^1S_0$ channel and in the $^3S_1$-$^3D_1$ coupled channels.
Assuming a rational, Bargmann-type S-matrix, a (\ref{one})-type small $r$
asymptotic behaviour naturally emerges. In this method the incompleteness of
the scattering data is compensated by the assumption on the rational form of
the S-matrix. For Bargmann-type S-matrices the strength parameter $\nu$ can
only take integer values. By making comparisons to phenomenological potentials
we found that $\nu=2$ in the $^1S_0$ channel and $\nu=1$, $\nu=3$ correspond
to the $^3S_1$ and $^3D_1$ channels, respectively (up to small mixing effects).

However, on physical grounds, there is no reason why the effective strength
parameter $\nu$ should be integer. In this paper we undertake a systematic
study of the strength parameter $\nu$ in various $np$ scattering channels
assuming the form (\ref{one}) but not requiring $\nu$ integer. We use the
Marchenko method of quantum inverse scattering because this efficient method
is applicable to all type of potentials (not necessarily of Bargmann-type).
In case of Bargmann potentials the Marchenko method has the extra advantage that
the results can be obtained purely algebraically \cite{EZB2}; in other cases
it requires the solution of a linear integral equation.

Quantum inverse scattering, the problem of finding the potential from
scattering data, is completely solved in the one-dimensional
case \cite{QIS1a, QIS1b, QIS1c} in a mathematically precise way.
The same mathematical problem emerges for three-dimensional spherically
symmetric potentials after partial wave expansion.
The potential can be uniquely reconstructed, in a given class of potentials,
if full information on scattering at all energies and some additional data
related to bound states (binding energies and asymptotic decay constants)
are all available. Since this is rarely the case, the method has not been
used often \cite{Selg} in nuclear theory, except in the case of Bargmann-type
potentials \cite{vonGeramb1, vonGeramb2}.

First, using the methods of \cite{QIS2}, we slightly generalized existing
results to incorporate singular potentials. Secondly, we worked out a method
to extrapolate limited range data so that the resulting potential is of
the form (\ref{one}). This is possible because the asymptotic large energy
behaviour is intimately related to the singularity strength via the
generalized Levinson's theorem.

We tested our method on an exactly solvable generalized P\"oschl-Teller
(Bargmann-type) potential, a slight generalization of those studied
in \cite{PT}. We found that the correct $\nu$ can be
reproduced with reasonable precision with our method.

Finally we undertook a systematic study of the $\nu$ parameter for various
low angular momentum partial waves of $np$ scattering: in the
$^1S_0$, $^3P_1$, $^3P_0$ and $^3D_2$ channels.
We constructed the potential in each channel with a
(\ref{one}) type singular behaviour based on experimental data below 350 MeV
LAB energy and extrapolated with some fixed $\nu$. We decided to compare
the resulting potential to the Reid93 phenomenological potential \cite{NuPo2}
at each channel, since it is a better description \cite{Machleidt} of
post-1993 data than the alternative AV18 \cite{NuPo3} phenomenological
potential. We determined the best choice for $\nu$ by requiring it gives the
best fit in the energy range 500-1000 MeV, somewhat above the validity range
of the original experimental data. We found that the singularity of the
central potential in the $^1S_0$ channel is still best approximated by
$\nu=2.0$, an integer.
(Visible deviations appear if we choose $\nu=1.8$ or $\nu=2.2$.)
But the best choice for the $^3P_1$, $^3P_0$, $^3D_2$ channels turn out to
be $\nu=2.3$, $\nu=3.2$, $\nu=2.3$, respectively.

As discussed above, the strength of the repulsive core (and the precise 
shape of the attraction pocket) depends on the particular choice of the 
phenomenological potential and this leads to a corresponding ambiguity 
in the determination of our strength parameter $\nu$. Since our 
favourite choice (the Reid93 parametrization) is just one particular 
choice, in the case of some channels (where both are available) we made 
a study of  $\nu$ values obtained by comparison to the AV18 
parametrization. We found that the optimal $\nu$ values differ only 
slightly between the two determinations.

The paper is organized as follows. In the next section we summarize Marchenko's
quantum inverse scattering algorithm. In section 3 we present our extrapolation
method for singular potentials, which is then applied in the next two sections
to $np$ scattering in various low angular momentum partial waves. Our
conclusions are summarized in section 6.


\section{Marchenko method of inverse scattering}
\label{marmeth}

In this paper we will apply the Marchenko method of inverse scattering
to $np$ scattering problems. Since our main focus is the singularity
strength of the repulsive core of the potential, we have generalized
the theory of quantum inverse scattering to the case of singular potentials
(see \cite{EZB2}). 

Our starting point is the radial Schr\"{o}dinger equation
\beq
\label{radialse}
-\frac{\hbar^2}{2m}u^{\prime\prime}(r)+\frac{\hbar^2}{2m}\,
\frac{\ell(\ell+1)} {r^2}u(r)+V(r)u(r)=Eu(r),
\eeq
where $m$ is the reduced mass\footnote{$m=469.459\,{\rm MeV}/c^2$ for the
$np$ system.}, $V$ is the interaction potential and
$E$ is the total energy of the particles. We introduce
\beq
q(r)=\frac{2m}{\hbar^2}V(r),\qquad k^2=\frac{2mE}{\hbar^2},
\qquad U(r)=q(r)+\frac{\ell(\ell+1)}{r^2},
\eeq
 which allows us to simplify (\ref{radialse}) as
\beq
-u^{\prime\prime}(r)+U(r)u(r)=k^2u(r).
\label{diff}
\eeq
 
We will consider potentials which are singular at the origin and vanish
exponentially at large distances,
\begin{eqnarray}
q(r)&\sim&\frac{\beta_o}{r^2}+{\rm O}(1), \qquad\quad \quad r\to0,\\
q(r)&\sim&e^{-2\kappa r},\qquad\quad r\to\infty, \quad (\kappa>0).  
\end{eqnarray}

The small $r$ singularity of the total potential term in (\ref{diff})
is determined by the parameter $\nu$ defined by
\beq
\nu(\nu+1)=\beta_o+\ell(\ell+1).
\label{beta0}
\eeq

Quantum inverse scattering is a method to reconstruct the potential
from scattering data. The latter is the set
\beq
\big\{S(k),0<k<\infty;\{s_j,\kappa_j\}_{j=1}^J\big\},
\label{Sdata}
\eeq
where $S(k)=\exp\{2i\delta(k)\}$ is the S-matrix. We see that we not
only need the phase shift $\delta(k)$ for all energies (all momenta $k$),
but also additional bound state information $\{s_j,\kappa_j\}$ for
all bound states $j=1,\dots,J$. Here $\kappa_j$ is related to the
binding energy by 
\beq
E_j=-\frac{\hbar^2\kappa_j^2}{2m}
\eeq
and $s_j$ to the asymptotic decay constant of the normalized bound
state wave function $\psi_j(r)$ by
\beq
s_j=A_j^2,\qquad\qquad \psi_j(r)\sim A_j{\rm e}^{-\kappa_j r},\qquad r\to\infty.
\eeq
The set of scattering data is only restricted by the requirement
\beq
\delta(k)={\rm O}(k^{2\ell+1}),\qquad k\to0
\eeq
(in the convention $\delta(0)=0$), and that the singularity strength $\nu$,
calculated from the generalized Levinsons's theorem \cite{EZB2},
\beq
\label{gLt}
\nu=\frac{2}{\pi}[\delta(0)-\delta(\infty)]-2J+\ell,
\eeq
must be non-negative (repulsive core).

If all scattering data are available, we first have to construct Marchenko's
$F(r,s)$ function, which is
the sum of a bound state contribution and a scattering contribution.
\beq
F(r,s)=F_{\rm bound}(r,s)+F_{\rm scatt}(r,s),
\label{Fbs}
\eeq 
where
\beq
F_{\rm bound}(r,s)=(-1)^{\ell+1}\sum_{j=1}^J s_j\
h_\ell(i\kappa_jr)\ h_\ell (i\kappa_js),
\label{Fb}
\eeq
and
\beq
\begin{split}
&F_{\rm scatt}(r,s)=-\frac{\sin\nu\pi}{\pi(r+s)}+\frac{1}{2\pi}
\int_0^\infty{\rm d}k\,\Big\{[S(k)-1]h_\ell(kr)h_\ell(ks)
+[S(-k)-1]h_\ell(-kr)h_\ell(-ks)\\ &+[1-S(\infty)](-1)^{\ell+1}\,{\rm e}^{ik(r+s)}
+[1-S(-\infty)](-1)^{\ell+1}\,{\rm e}^{-ik(r+s)}\Big\}.
\end{split}
\label{Fs}
\eeq
Here $h_\ell(x)$ is the Riccati-Hankel function, defined as
\beq
h_\ell(x)=(-i)^{\ell+1}e^{ix}\sum^\ell_{m=0}\frac{i^m}{m!(2x)^m}
\frac{(\ell+m)!}{(\ell-m)!}.
\label{han}
\eeq
The scattering part can be rewritten after partial integration as
\beq
F_{\rm scatt}(r,s)=\frac{2}{\pi}(-1)^\ell\frac{1}{r+s}\int_0^\infty{\rm d}k\,
\delta^\prime(k){\rm Re}\,\left\{S(k){\rm e}^{ik(r+s)}R_\ell(r,s,k)\right\},
\label{FRell}
\eeq  
where
\beq
R_\ell(r,s,k)=\frac{1}{s-r}\left[sP_\ell(kr)P_{\ell-1}(ks)
-rP_\ell(ks)P_{\ell-1}(kr)\right]
\eeq
and $P_\ell$ is the polynomial part of the Hankel function (\ref{han}):
\beq
P_\ell(x)=\sum_{m=0}^\ell C_{\ell,m}\left(\frac{i}{x}\right)^m,\qquad\quad
C_{\ell,m}=\frac{(\ell+m)!}{2^m m!(\ell-m)!}.
\eeq

After some algebra, we can write
\beq
R_\ell(r,s,k)=\sum_{m=0}^\ell C_{\ell,m}\frac{Z_m(-i(r+s)k)}{(-rsk^2)^m},
\label{Rell}
\eeq
where $Z_m(\xi)$ is a polynomial of degree $m$. The first few
polynomials are
\beq
Z_0(\xi)=1,\qquad Z_1(\xi)=\xi,\qquad Z_2(\xi)=\xi(\xi+1),\qquad
Z_3=\xi(\xi^2+3\xi+3)
\eeq
and
\beq
Z_4=\xi(\xi^3+6\xi^2+15\xi+15).
\eeq
This gives
\begin{eqnarray}
R_0(r,s,k)&=&1,\label{R0}\\
R_1(r,s,k)&=&1+\frac{i(r+s)}{rsk},\label{R1}\\
R_2(r,s,k)&=&1+\frac{3i(r+s)}{rsk}-\frac{3i(r+s)}{(rs)^2k^3}
-\frac{3(r+s)^2}{(rs)^2k^2},\label{R2}
\end{eqnarray}
etc. (\ref{Rell}) is most convenient for numerical integration because
it reduces the calculation of Marchenko's $F$, a function of two
variables, to the calculation of $\ell+1$ functions of one variable.

The next step is to solve the Marchenko equation
\beq
F(r,s)+A(r,s)+\int_r^\infty A(r,u) F(u,s)\, {\rm d}u=0,\qquad\quad s\geq r
\label{mareq}
\eeq
for $A(r,s)$. Finally one has to take the derivative of $A(r,r)$
and obtain the potential in the Schr\"odinger equation (\ref{diff}) by
\beq
q(r)=-2\frac{{\rm d}}{{\rm d}r}\,A(r,r),
\quad \qquad U(r)=q(r)+\frac{\ell(\ell+1)}{r^2}.
\eeq


\section{Singular potentials and incomplete data}
\label{inc}

Marchenko's method is useful only if we have access to the full set of
scattering data (\ref{Sdata}). In our case of the nonrelativistic nucleon
potential, as explained in the introduction, for various reasons we can
only use low energy data up to about 350~MeV LAB energy. Since for the
uncoupled channels we are considering there are no bound states, the only
missing piece is scattering phase shifts for energies above the maximal
energy, which we take to be 350 MeV.

In this paper we adopted the following strategy. We use measured phase
shifts up to the maximal energy and smoothly extrapolate this function
for higher energies taking into account the singularity strength of the
potential, which, in the absence of bound states, is related to the
asymptotic value of the phase shift by the relation
\beq
\delta(\infty)=-\frac{\pi}{2}(\nu-\ell).
\eeq
In more detail, we use as our phase shift
\beq
\delta(k)=\left\{\begin{split}
&\delta_{\rm int}(k),\qquad 0<k<k_o,\\
&\delta_{\rm ext}(k),\qquad k>k_o,
\end{split}\right.
\eeq
where $\delta_{\rm int}$ is an interpolating function based on the measured
data points with the only constraint
\beq
\delta_{\rm int}(k)={\rm O}(k^{2\ell+1})
\eeq
for small $k$, and $\delta_{\rm ext}$ is the extrapolated part with
\beq
\delta_{\rm ext}(\infty)=-\frac{\pi}{2}(\nu-\ell)
\eeq
asymptotics. Moreover, at the point $k=k_o=2.054\, {\rm fm}^{-1}$, which
corresponds to the maximal energy, we require that the interpolation
and the extrapolation are joined smoothly. We made the simple choice
\beq
\delta_{ext}(k)=-\frac{\pi}{2}(\nu-\ell)+\sum_{m=1}^3\frac{a_m}{k^m}
\label{extrapol}
\eeq
and determined the three constants $a_m$ from the requirements that the value,
the slope and the curvature of the two functions coincide at $k=k_o$.

In the next subsection we study a test example, which shows that this
simple method work reasonably well, at least in the class of functions
resembling the phenomenological nucleon potential.


\subsection{A test example}
\label{test}

To test our method we have chosen the $\ell=0$ phase shift
\beq
\delta_{\rm exact}(k)=-\pi +\sum_{n=1}^6 \arctan \left(\frac{x_n}{k}\right),
\label{delex}
\eeq
where
\beq
x_1=-0.0401, 
x_2=-0.7540,   
x_3=0.6152,   
x_4=2.0424, 
x_5=4.1650,
x_6=4.6000. \quad
\eeq

\begin{figure}
        \centering
        \begin{subfigure}[b]{0.31\textwidth}
                \centering
                \includegraphics[width=\textwidth]{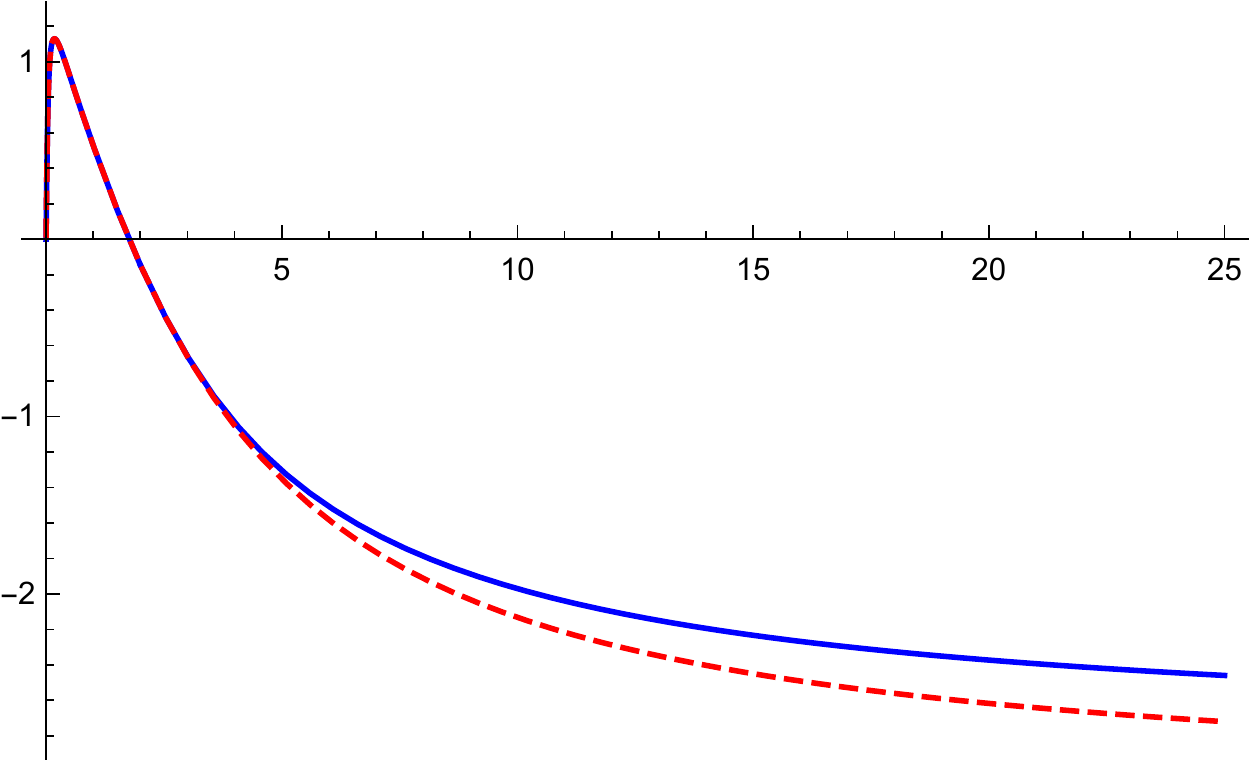}
                \caption{$\nu=1.8$}
        \end{subfigure}%
        ~ 
        \begin{subfigure}[b]{0.31\textwidth}
                \centering
                \includegraphics[width=\textwidth]{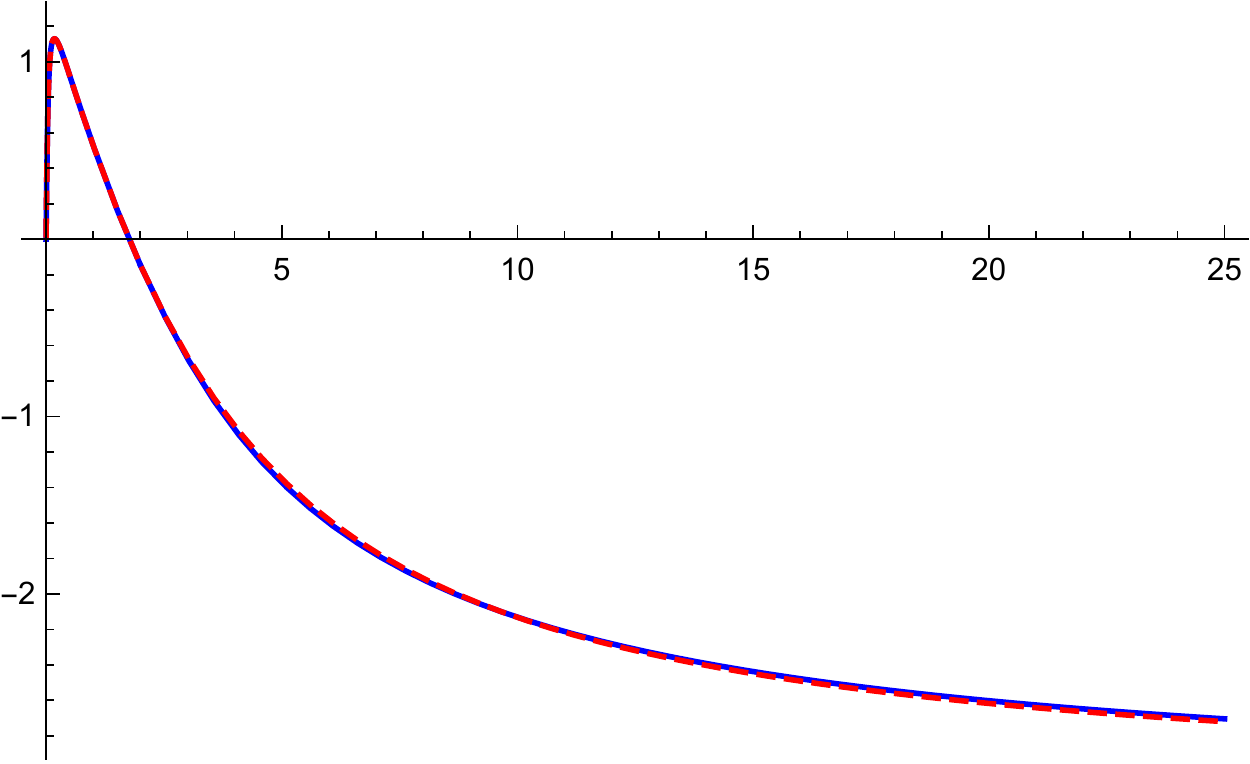}
                \caption{$\nu=2.0$}
        \end{subfigure}
        ~ 
        \begin{subfigure}[b]{0.31\textwidth}
                \centering
                \includegraphics[width=\textwidth]{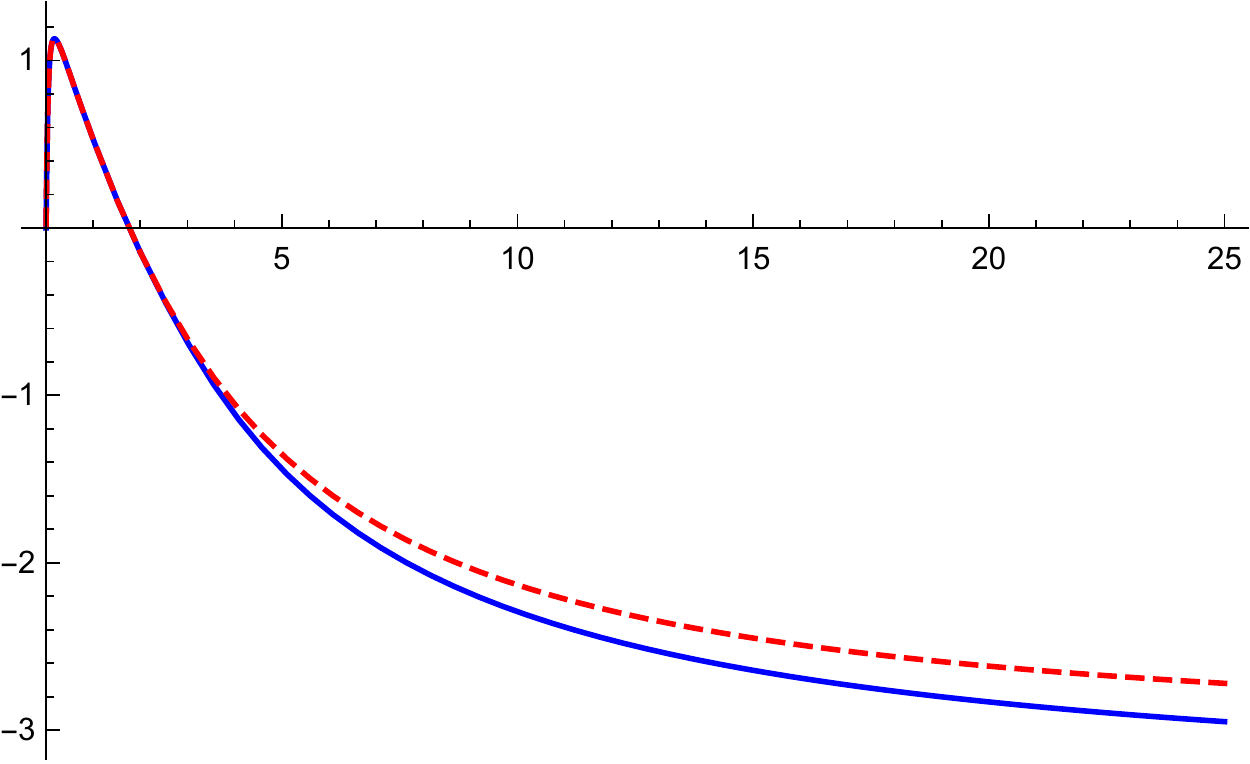}
                \caption{$\nu=2.2$}
        \end{subfigure}
        \caption{\footnotesize{Reconstruction of the phase shift $\delta(k)$ (radian)
                  with different singularity strength values. The solid
                  (blue) line is the phase shift obtained with our
                  extrapolation method and the dashed (red) line is the
                  original phase shift. The extrapolation starts at
                  $k=2$.}}
\label{del25test}
\end{figure}

\begin{figure}
        \centering
        \begin{subfigure}[b]{0.45\textwidth}
                \centering
                \includegraphics[width=\textwidth]{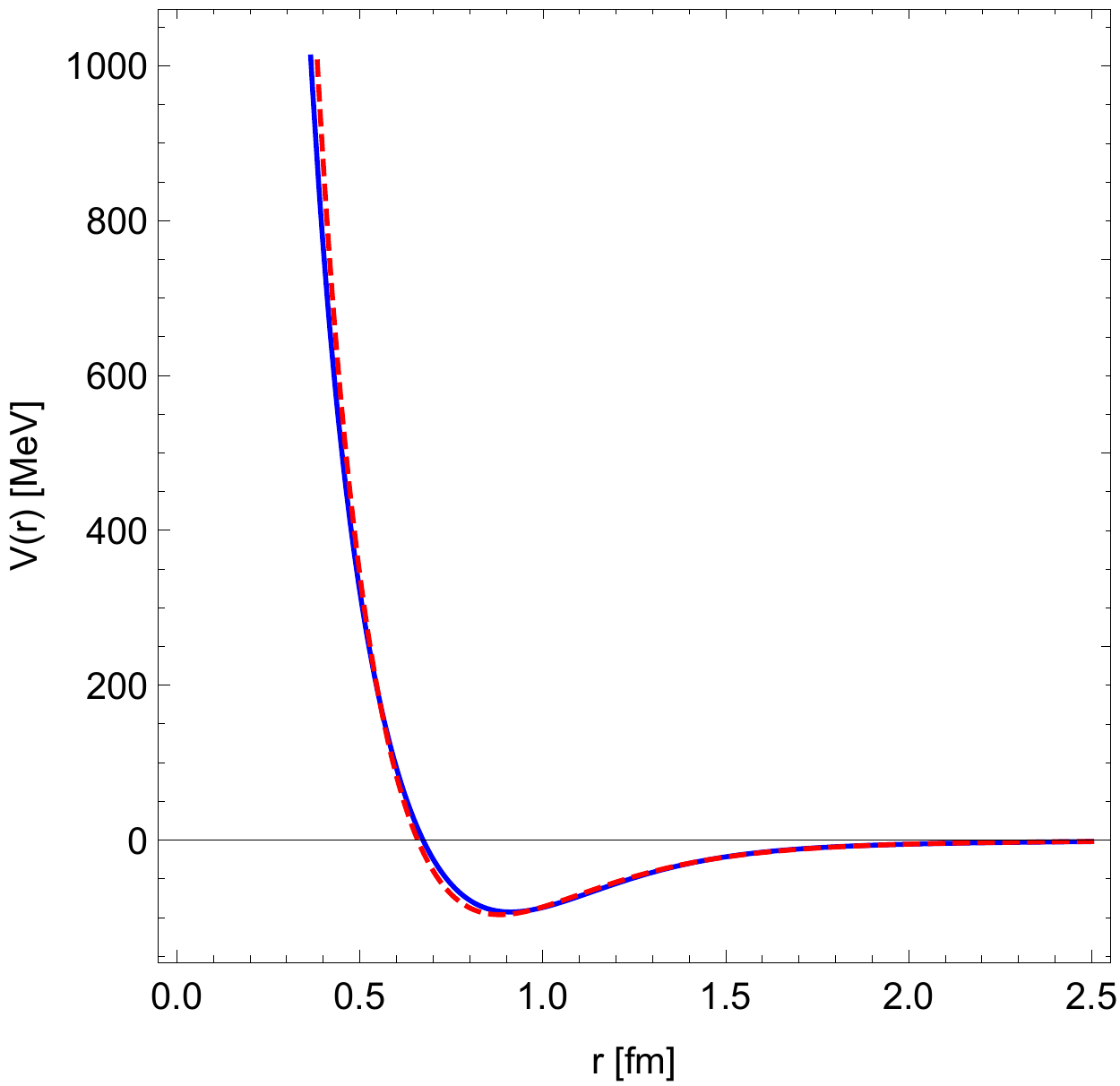}
                \caption{$\nu=1.8$}
        \end{subfigure}%
        ~ 
        \begin{subfigure}[b]{0.45\textwidth}
                \centering
                \includegraphics[width=\textwidth]{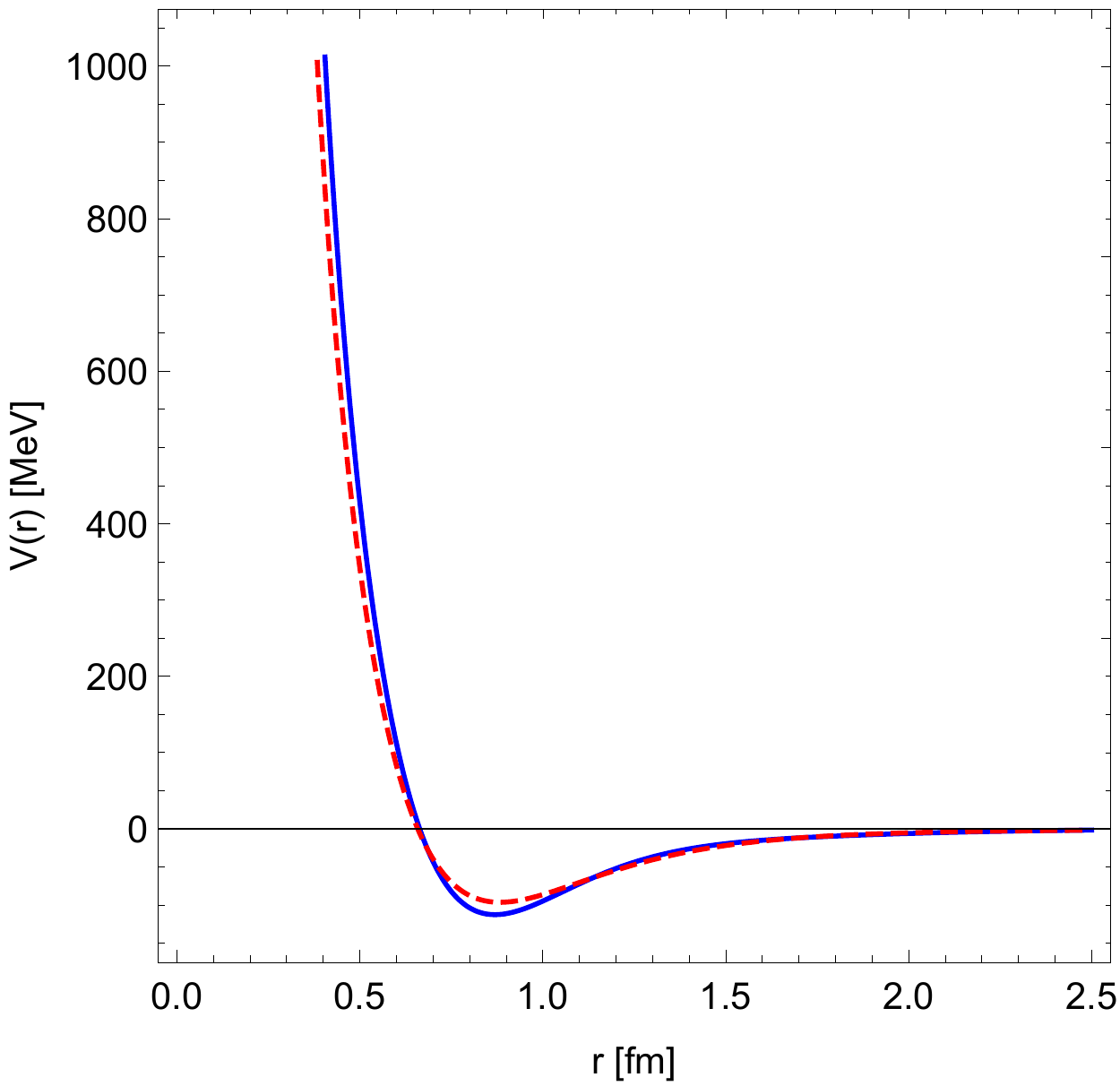}
                \caption{$\nu=2.2$}
        \end{subfigure}
        ~ 
        \begin{subfigure}[b]{0.45\textwidth}
                \centering
                \includegraphics[width=\textwidth]{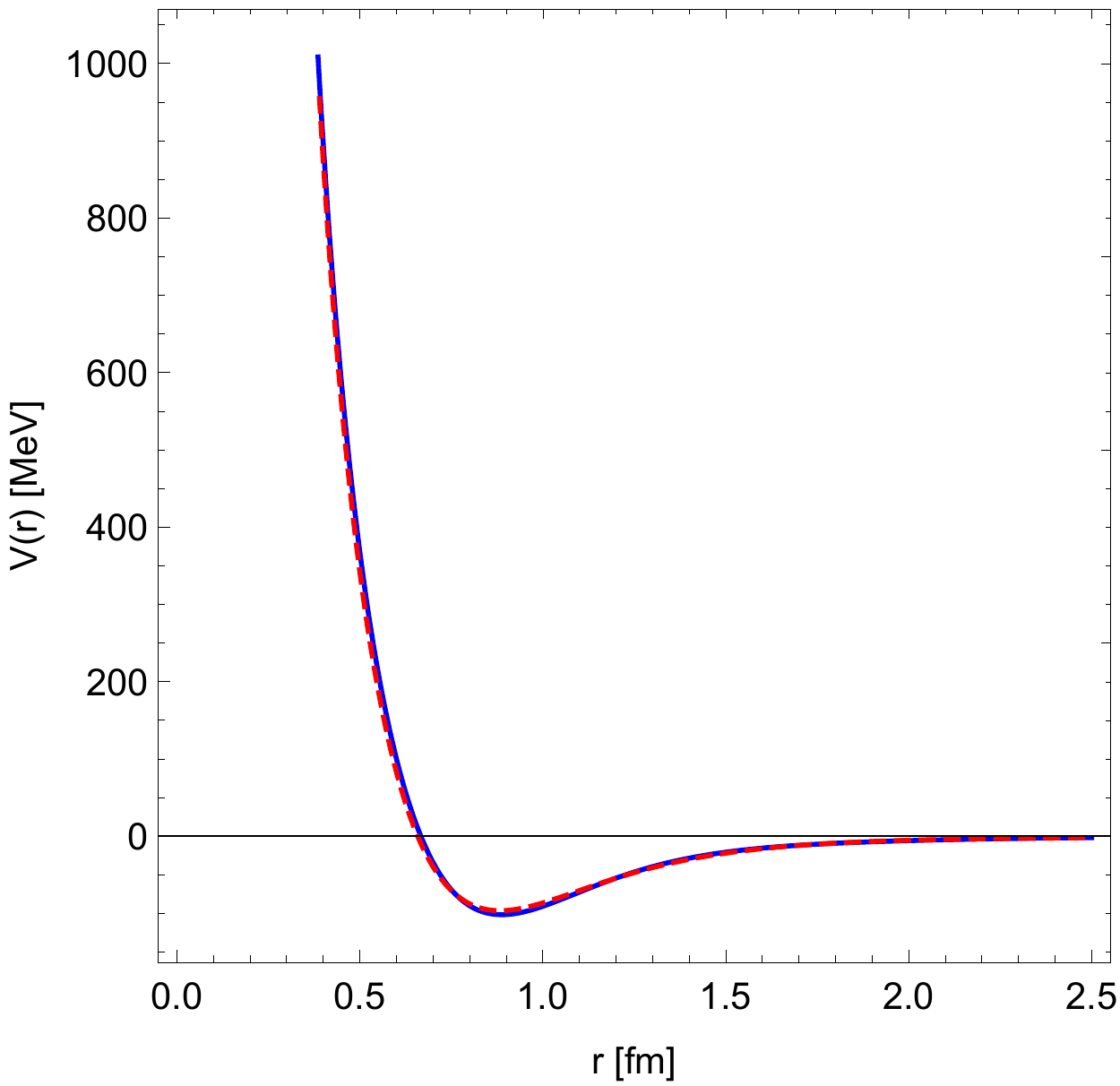}
                \caption{$\nu=2.0$}
        \end{subfigure}
        \caption{\footnotesize{Reconstruction of the test potential with different
                  singularity strength values. The solid (blue) line
                  is the potential obtained with our extrapolation
                  method and the dashed (red) line is the original test
                  potential.}}
\label{pottest}
\end{figure}

Assuming the absence of bound states, from Levinson's theorem 
we see that this phase shift corresponds to a Bargmann-type S-matrix with
$\nu=2$. The parameter values are from \cite{Stancu} and we also used this
Bargmann S-matrix in \cite{EZB2} to represent the phenomenological potential
for $np$ scattering in the $^1S_0$ channel. However, for our present purposes
it just serves as an exactly known test potential, which is qualitatively
similar to the nucleon phenomenological potential.

To mimic what we want to do in the real case later, we approximate the
phase shift as
\beq
\delta(k)=\left\{\begin{split}
&\delta_{\rm exact}(k),\qquad 0<k<2,\\
&\delta_{\rm ext}(k),\qquad \qquad k>2,
\end{split}\right.
\eeq
where the extrapolation is of the form (\ref{extrapol}) and the two parts
are glued together at $k=2$ requiring that the values, slopes and curvatures
match at that point. 

We have considered the values $\nu=1.8$, $\nu=2.0$ and $\nu=2.2$. The
reconstruction of the potential for the original, Bargmann-type phase
shift (\ref{delex}) is of course already known \cite{Stancu, EZB2}.
For the extrapolated phase shifts we calculated both Marchenko's $F$
function and the solution of the Marchenko equation numerically.
The extrapolated phase shifts are compared to the original in
Fig.~\ref{del25test}, while the potential reconstructed
with the Marchenko method is shown in Fig.~\ref{pottest}. As we can see,
for $\nu=1.8$ and $\nu=2.2$ some deviation (with opposite sign) is clearly
visible, while for the true value $\nu=2.0$ the agreement of our extrapolated
phase shift and potential with the original is quite good. We can conclude
that our extrapolation method works reasonably well and the \lq\lq true''
value of the singularity strength $\nu$ can be estimated.


\section{$^1S_0$ channel as an example of inverse scattering with
incomplete data}
\label{1S0}

In this section we apply our extrapolation method to the central potential
in the $^1S_0$ isovector channel.

\begin{figure}
        \centering
        \begin{subfigure}[b]{0.45\textwidth}
                \centering
                \includegraphics[width=\textwidth]{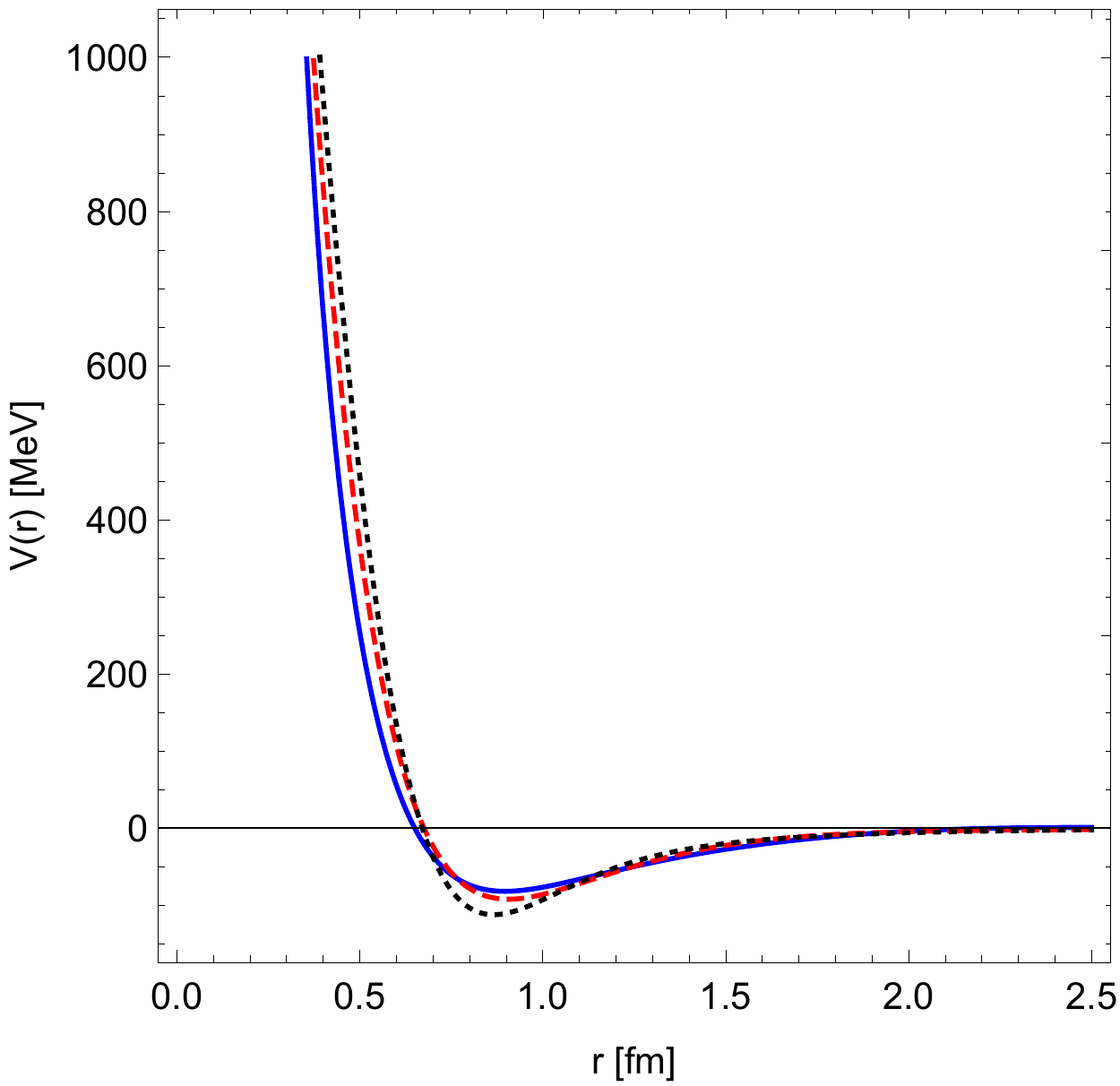}
                \caption{$\nu=1.8$}
        \end{subfigure}%
        ~ 
        \begin{subfigure}[b]{0.45\textwidth}
                \centering
                \includegraphics[width=\textwidth]{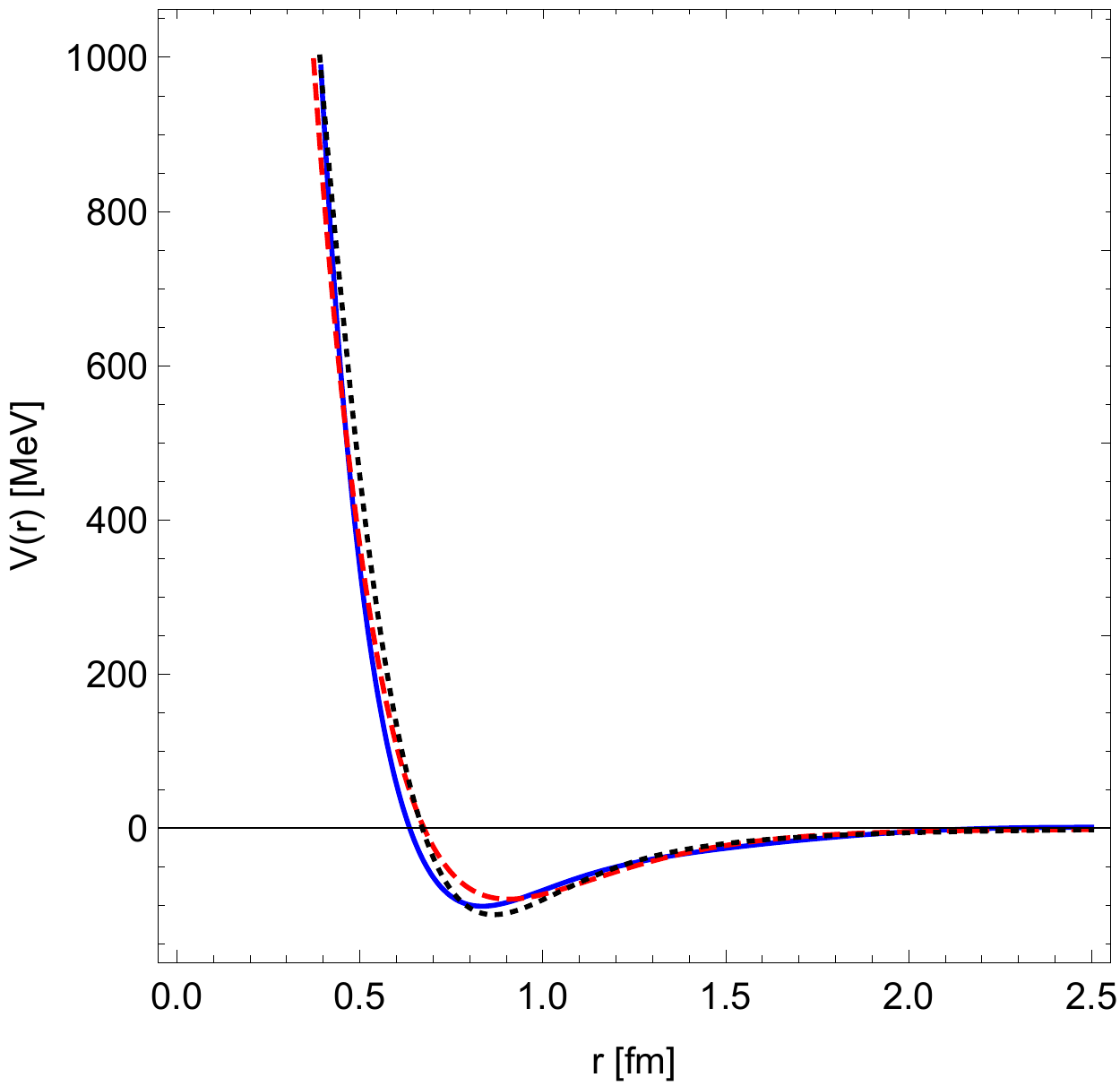}
                \caption{$\nu=2.2$}
        \end{subfigure}
        ~ 
        \begin{subfigure}[b]{0.45\textwidth}
                \centering
                \includegraphics[width=\textwidth]{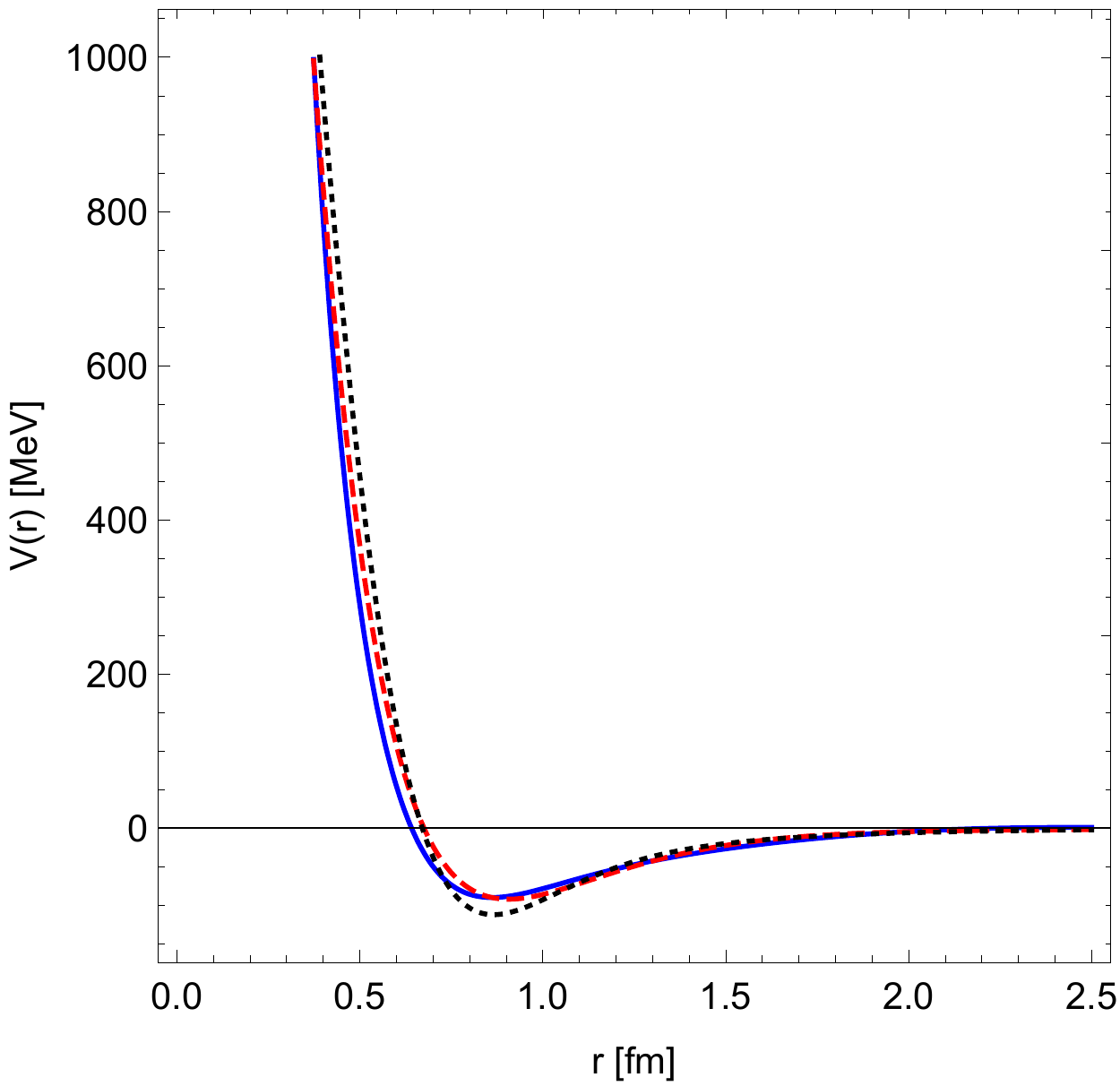}
                \caption{$\nu=2.0$}
        \end{subfigure}
        \caption{\footnotesize{Reconstruction of the $^1S_0$ channel central potential
          with different singularity strength values. The solid
          (blue) line is the potential obtained with our extrapolation
                  method and the dashed (red) line is the Reid93 \cite{NuPo2}
                  potential. For comparison, the AV18 potential \cite{NuPo3}
                  is also shown (dotted, black).}}
\label{pot1S0}
\end{figure}

For this purpose we downloaded the phase shifts from the publicly
available GWDAC data base \cite{DATA}. We use 35 phase shift data for LAB
energies between 0 and 350 MeV. For concreteness, we have chosen
the results of the analysis of Ref.~\cite{DATA1} (unweighted fit).
Of course, these results are already processed and not raw experimental data
but for simplicity we call them \lq experimental' data. The results of other
phase shift analyses differ very little from these, in the potential model
region we are studying.

Next we constructed our phase shift function by smoothly gluing together the
interpolation to the experimental data with the extrapolation for higher
energies as explained in section \ref{inc}. Because in our previous
study \cite{EZB2} based on Bargmann-type extrapolation we found that $\nu=2$
gives satisfactory results, we have considered the parameter values $\nu=1.8$,
$\nu=2.0$ and $\nu=2.2$ here.

Having constructed our $\delta(k)$, we calculated Marchenko's $F$ function
numerically using the formula (see (\ref{FRell}), (\ref{R0}))
\beq
F_0(r,s)=F_{(0)}(x)=\frac{2}{\pi x}
\int_0^\infty{\rm d}k\,\delta^\prime(k)\cos[2\delta(k)+kx],
\label{F0int}
\eeq
where $x=r+s$. There is no bound state contribution in this problem.

Finally we solved the Marchenko integral equation (\ref{mareq}) numerically.
The calculation is greatly simplified by noticing that for every fixed $r$
we have a separate problem, parametrized by $r$, where only the $s$ variable
is dynamical.

\begin{figure}
\begin{flushleft}
\vskip -30mm
\leavevmode
\centerline{\includegraphics[width=7cm]{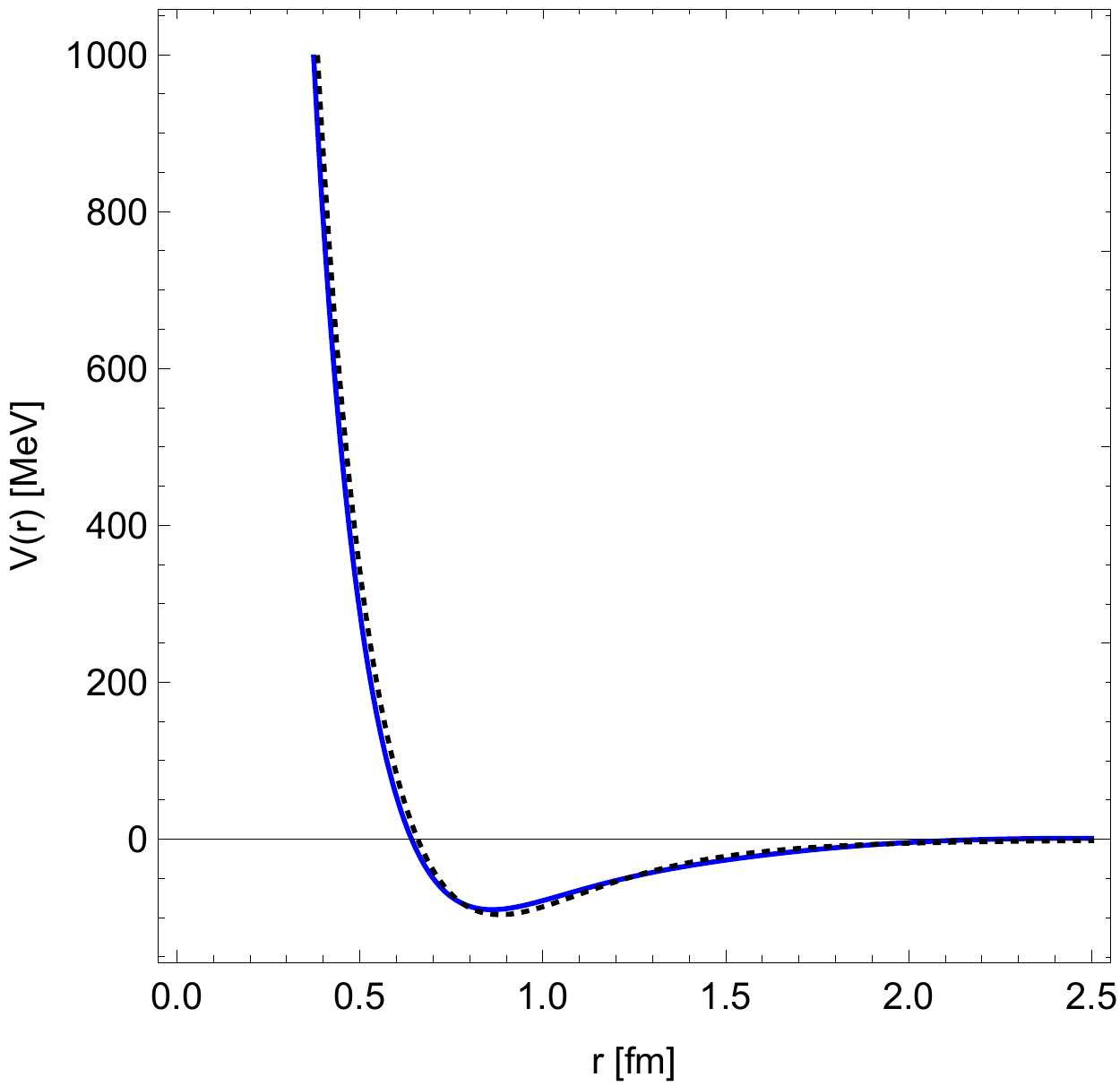} }
\end{flushleft}
\vskip -0.5cm
\caption{{\footnotesize
    Comparison of the reconstructed $^1S_0$ channel potential at $\nu=2.0$
    (solid, blue) to the Bargmann-type potential (dotted, black).
}}
\label{potBarg}
\end{figure}

The results are shown in Fig. \ref{pot1S0}. Here we compare our potentials,
reconstructed by the Marchenko method from our extrapolated phase shifts,
for the three different strength parameter values with the Reid93 \cite{NuPo2}
phenomenological potential. (In the plots the alternative AV18 \cite{NuPo3}
phenomenological potential is also shown.) The figure suggests that $\nu=2.0$
is the best choice here. For $\nu=1.8$ and $\nu=2.2$ there is already a
noticable deviation (in opposite directions) from Reid93 in the high energy
range above 500 MeV. Thus our conclusion is that the integer value $\nu=2$
found previously in the class of Bargmann potentials remains the best choice
here, even if we allow for non-integer values of the parameter. Actually, the
potential reconstructed by the above method and the Bargmann potential differ
by very little, as shown in Fig. \ref{potBarg}.
$\nu=2$ corresponds to $\beta_o=6$ in (\ref{beta0}).


\section{Other channels}
\label{other}

\subsection{The $^3P_1$, $^3P_0$ and $^3D_2$ channels}

We have also studied the potentials corresponding to the uncoupled channels
$^3P_1$, $^3P_0$ and $^3D_2$ with the same method.

\begin{figure}
        \centering
        \begin{subfigure}[b]{0.45\textwidth}
                \centering
                \includegraphics[width=\textwidth]{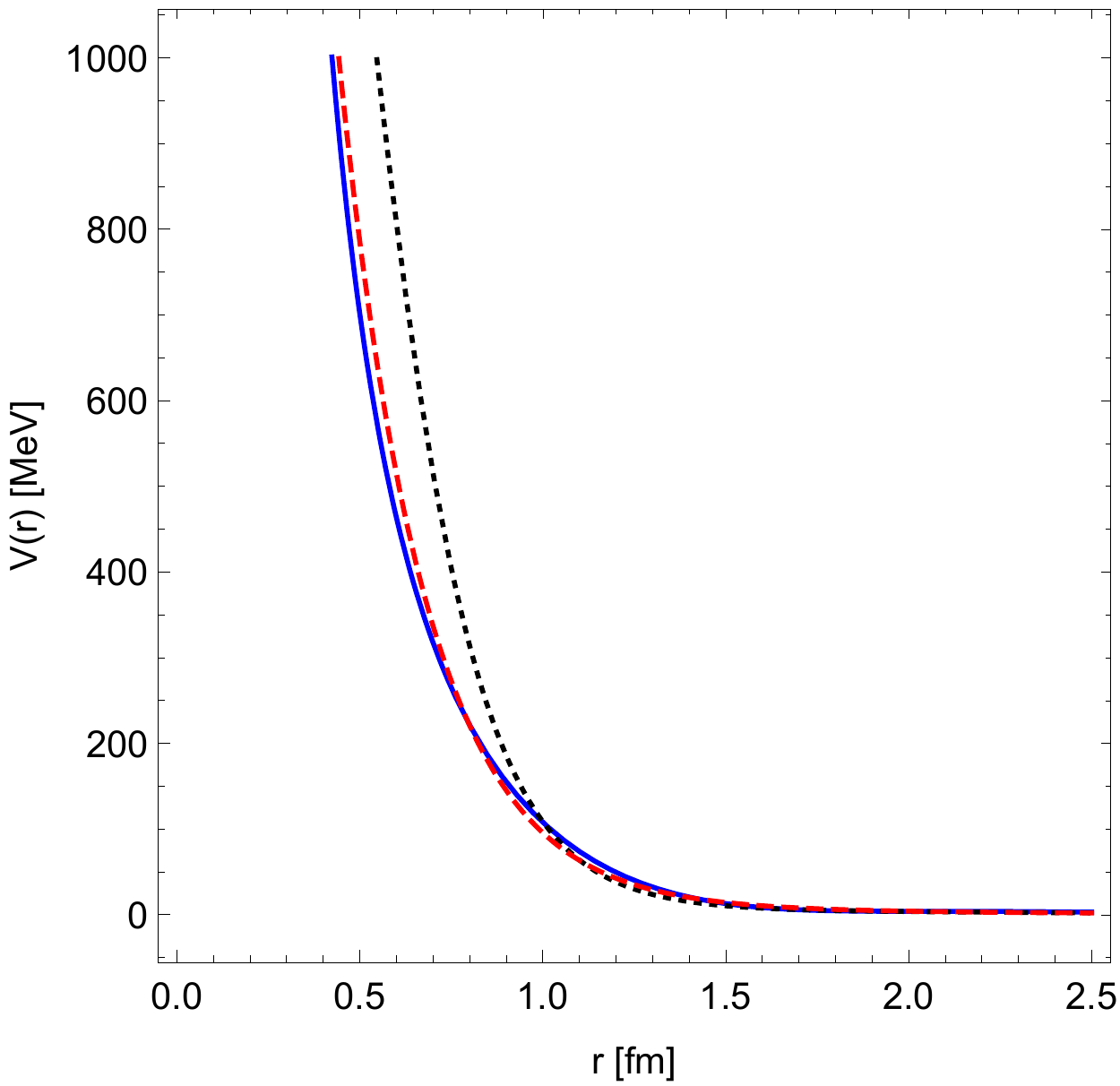}
                \caption{$\nu=2.1$}
        \end{subfigure}%
        ~ 
        \begin{subfigure}[b]{0.45\textwidth}
                \centering
                \includegraphics[width=\textwidth]{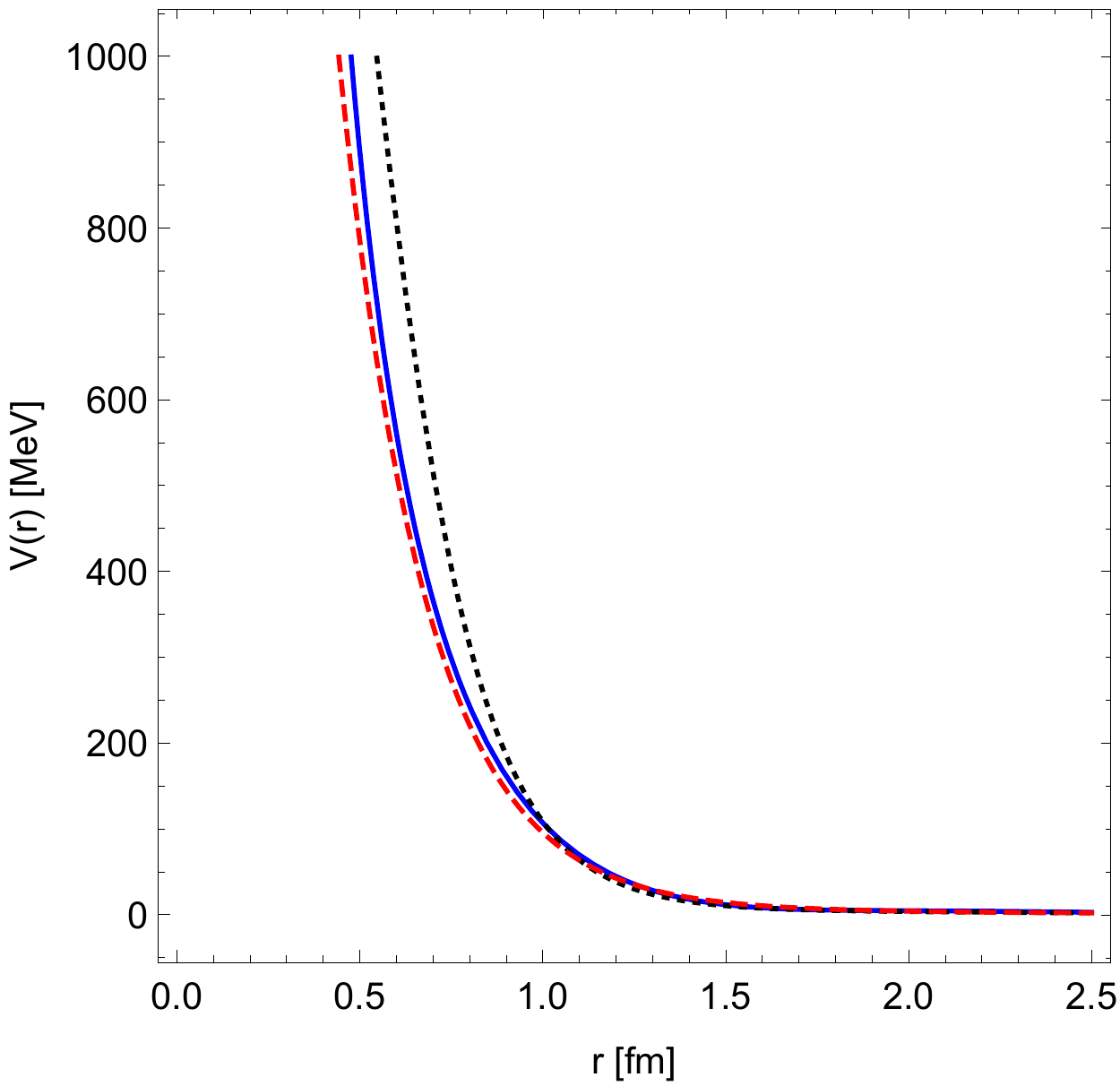}
                \caption{$\nu=2.5$}
        \end{subfigure}
        ~ 
        \begin{subfigure}[b]{0.45\textwidth}
                \centering
                \includegraphics[width=\textwidth]{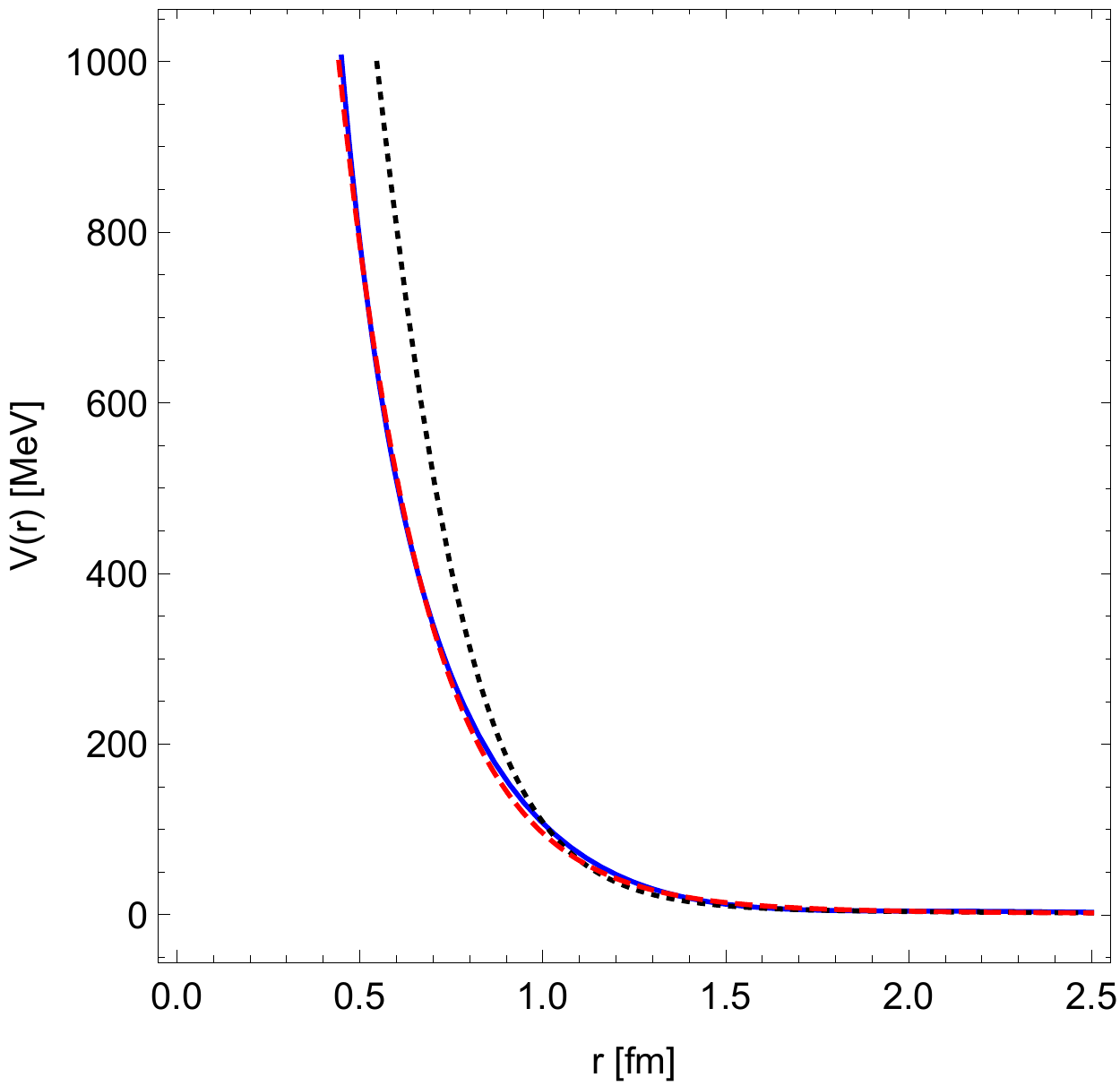}
                \caption{$\nu=2.3$}
        \end{subfigure}
        \caption{\footnotesize{Reconstruction of the $^3P_1$ channel nucleon potential
          with different singularity strength values. The solid
          (blue) line is the potential obtained with our extrapolation
                  method and the dashed (red) line is the Reid93 \cite{NuPo2}
                  potential. For comparison, the AV18 potential \cite{NuPo3}
                  is also shown (dotted, black).}}
\label{pot3P1}
\end{figure}

For $^3P_1$ and $^3P_0$ the angular momentum is $\ell=1$ and we have the
constraint that the interpolating function near $k=0$ has to be ${\rm O}(k^3)$.
Marchenko's $F$ function is given\footnote{Note that there are no bound states
in any of the channels discussed in this subsection.} by the
formula (see (\ref{FRell}), (\ref{R1}))
\beq
F_1(r,s)=-F_{(0)}(x)-\frac{F_{(1)}(x)}{rs},
\eeq
where $F_{(0)}$ is given by (\ref{F0int}),
\beq
F_{(1)}(x)=-\frac{2}{\pi}
\int_0^\infty\frac{{\rm d}k}{k}\delta^\prime(k)\sin[2\delta(k)+kx]
\label{F1int}
\eeq
and $x=r+s$.

\begin{figure}
        \centering
        \begin{subfigure}[b]{0.45\textwidth}
                \centering
                \includegraphics[width=\textwidth]{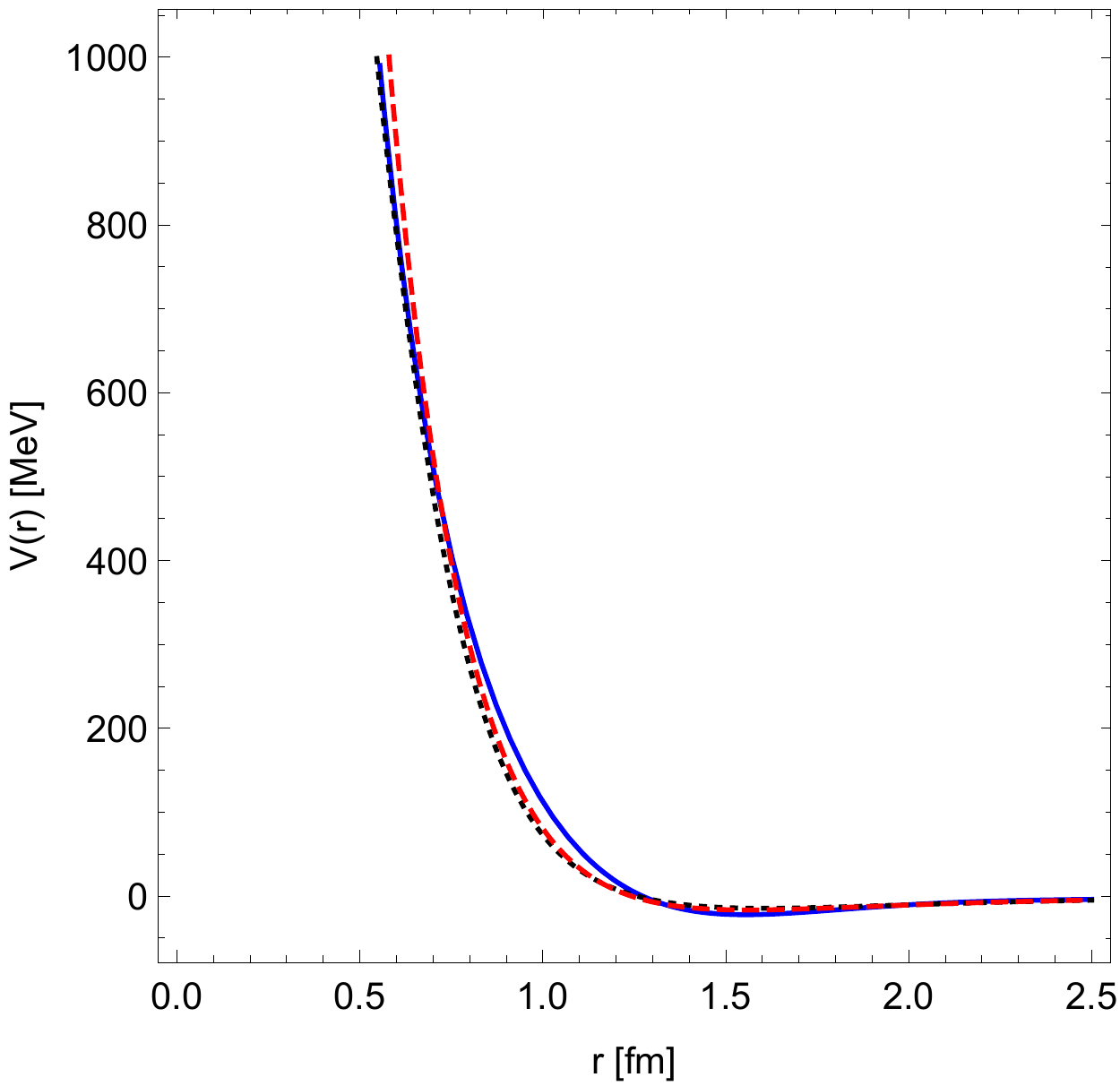}
                \caption{$\nu=3.0$}
        \end{subfigure}%
        ~ 
        \begin{subfigure}[b]{0.45\textwidth}
                \centering
                \includegraphics[width=\textwidth]{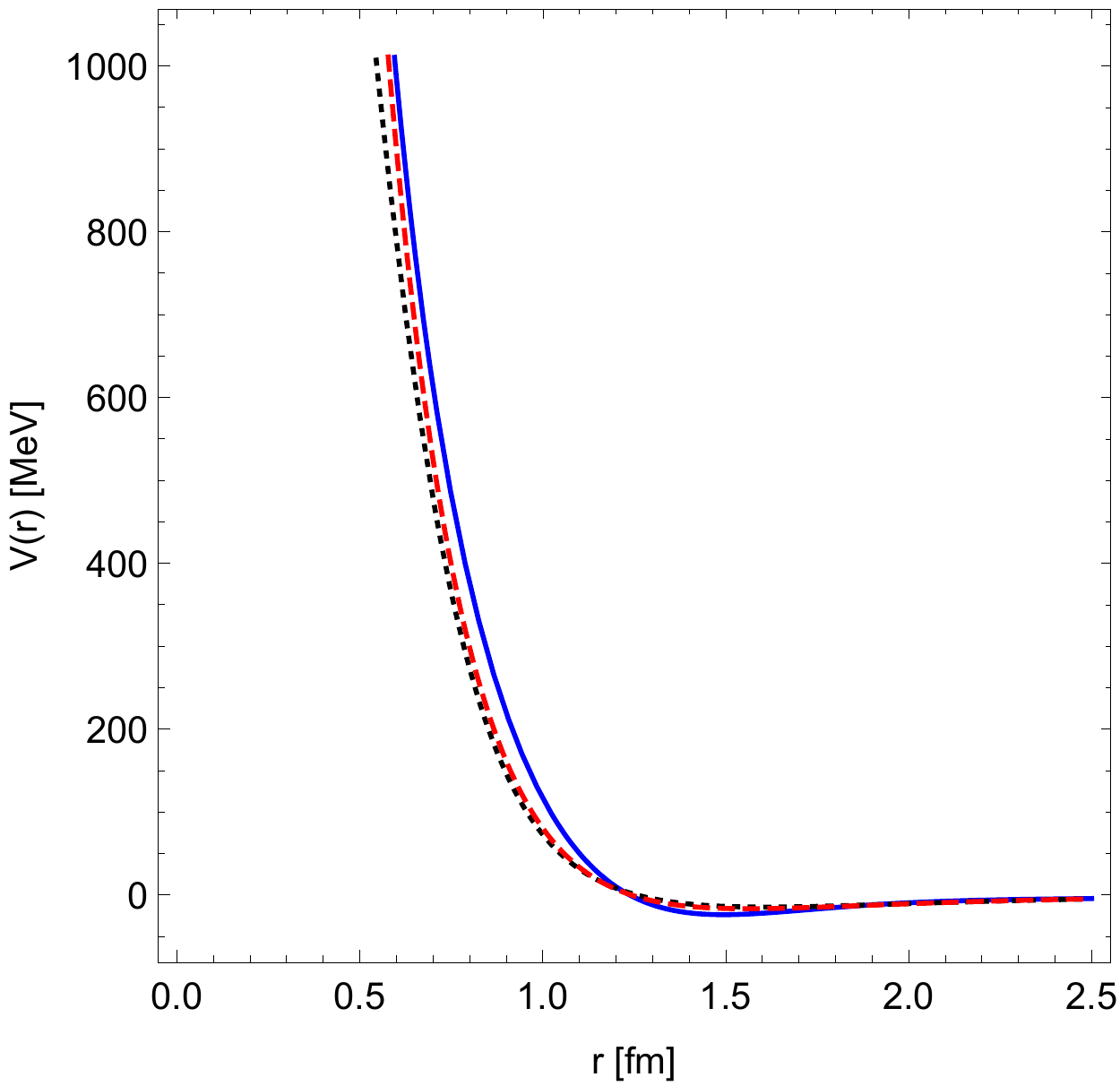}
                \caption{$\nu=3.4$}
        \end{subfigure}
        ~ 
        \begin{subfigure}[b]{0.45\textwidth}
                \centering
                \includegraphics[width=\textwidth]{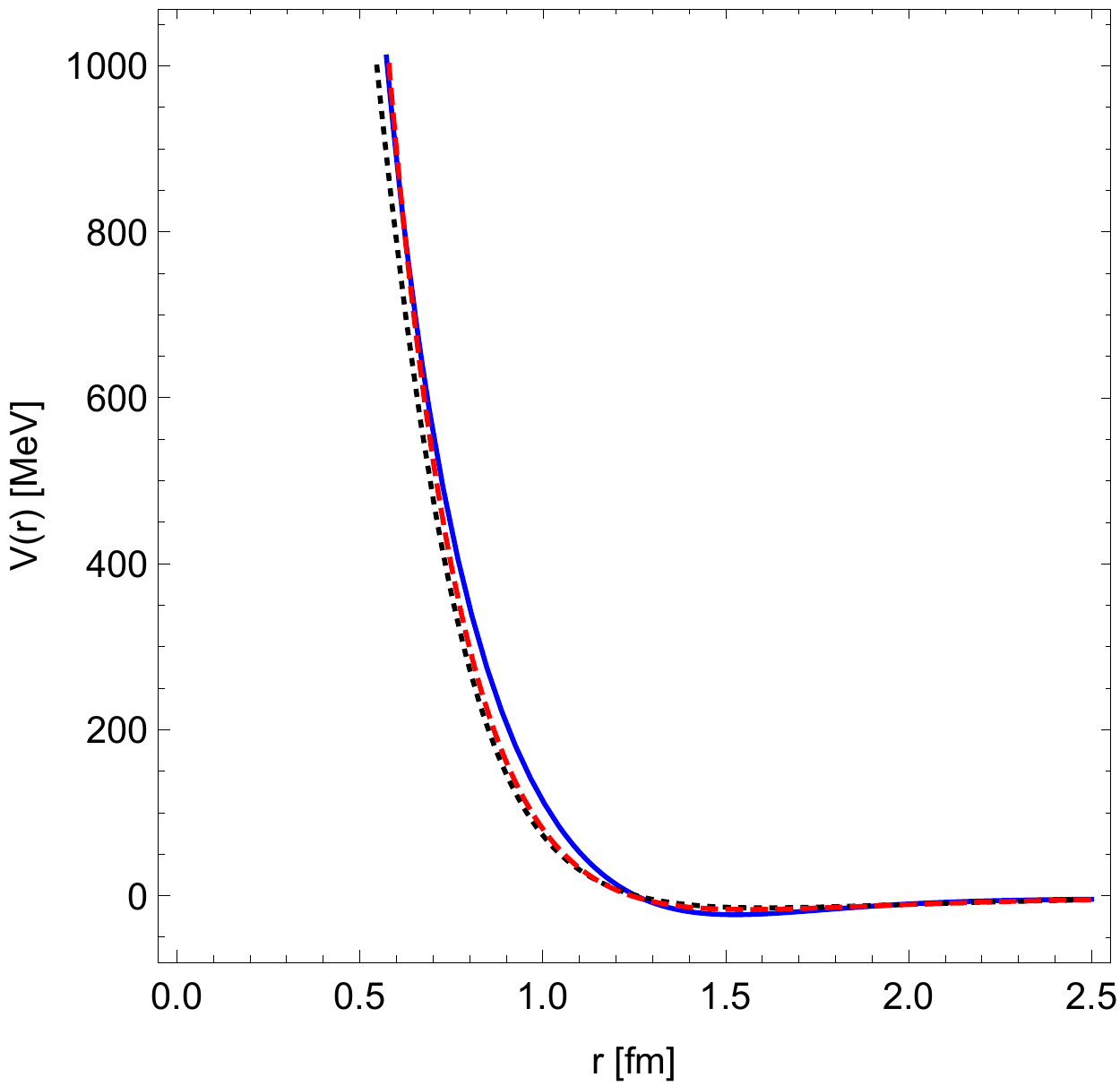}
                \caption{$\nu=3.2$}
        \end{subfigure}
        \caption{\footnotesize{Reconstruction of the $^3P_0$ channel nucleon potential
          with different singularity strength values. The solid
          (blue) line is the potential obtained with our extrapolation
                  method and the dashed (red) line is the Reid93 \cite{NuPo2}
                  potential. For comparison, the AV18 potential \cite{NuPo3}
                  is also shown (dotted, black).}}
\label{pot3P0}
\end{figure}

For $^3D_2$ $\ell=2$ and the constraint\footnote{The constraints are necessary
to make the integrals (\ref{F1int},\ref{F2int}) convergent at $k=0$.}
is $\delta(k)={\rm O}(k^5)$.
Marchenko's $F$ function is given by (see (\ref{FRell}), (\ref{R2}))
\beq
F_2(r,s)=F_{(0)}(x)+3\frac{F_{(1)}(x)}{rs}+3\frac{F_{(2)}(x)}{(rs)^2},
\eeq
where $F_{(0)}$ is given by (\ref{F0int}), $F_{(1)}$ by (\ref{F1int}) and
\beq
F_{(2)}(x)=\frac{2}{\pi}
\int_0^\infty{\rm d}k\,\delta^\prime(k)\Big\{
\frac{\sin[2\delta(k)+kx]}{k^3}-\frac{x\cos[2\delta(k)+kx]}{k^2}\Big\}.
\label{F2int}
\eeq

\begin{table}[t]
\centering
\begin{tabular}[t]{c|c|c|c}
\hline
channel & $\ell$ value & best fit $\nu$ value & $\beta_o $ value 
\\[0.5ex]
\hline \hline
$^1S_0$ & 0 & 2.0 &
6.0 \\[0.5ex]
$^3P_1$ & 1 & 2.3 &
5.6 \\[0.5ex]
$^3P_0$ & 1 & 3.2 &
11.4 \\[0.5ex]
$^3D_2$ & 2 & 2.3 &
1.6 \\[0.5ex]
\hline
\end{tabular}
\caption{\footnotesize Fractional $\nu$ values and the corresponding $\beta_o$ values }
\label{Tab1}
\end{table}

\begin{figure}
        \centering
        \begin{subfigure}[b]{0.45\textwidth}
                \centering
                \includegraphics[width=\textwidth]{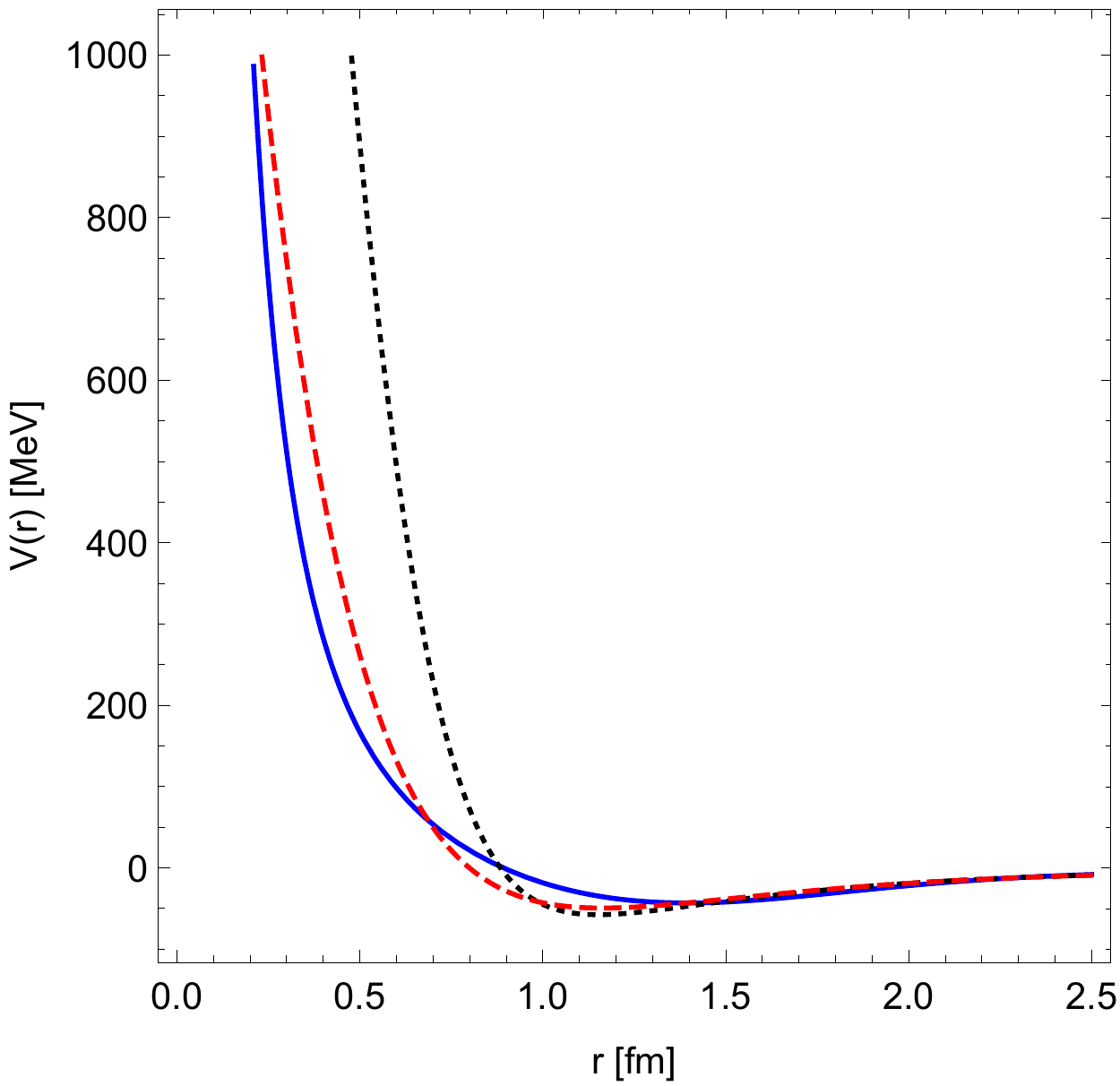}
                \caption{$\nu=2.2$}
        \end{subfigure}%
        ~ 
        \begin{subfigure}[b]{0.45\textwidth}
                \centering
                \includegraphics[width=\textwidth]{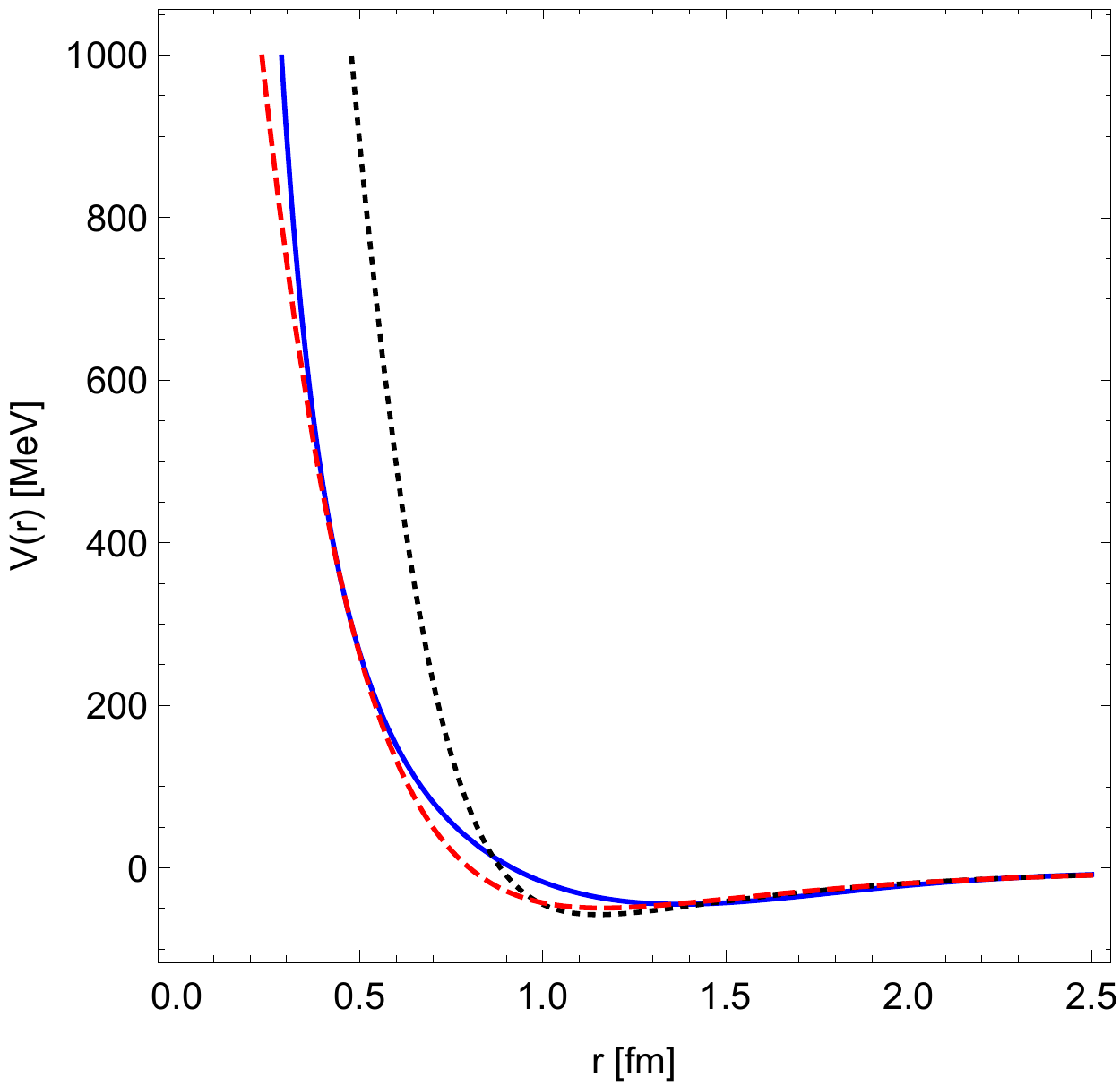}
                \caption{$\nu=2.4$}
        \end{subfigure}
        ~ 
        \begin{subfigure}[b]{0.45\textwidth}
                \centering
                \includegraphics[width=\textwidth]{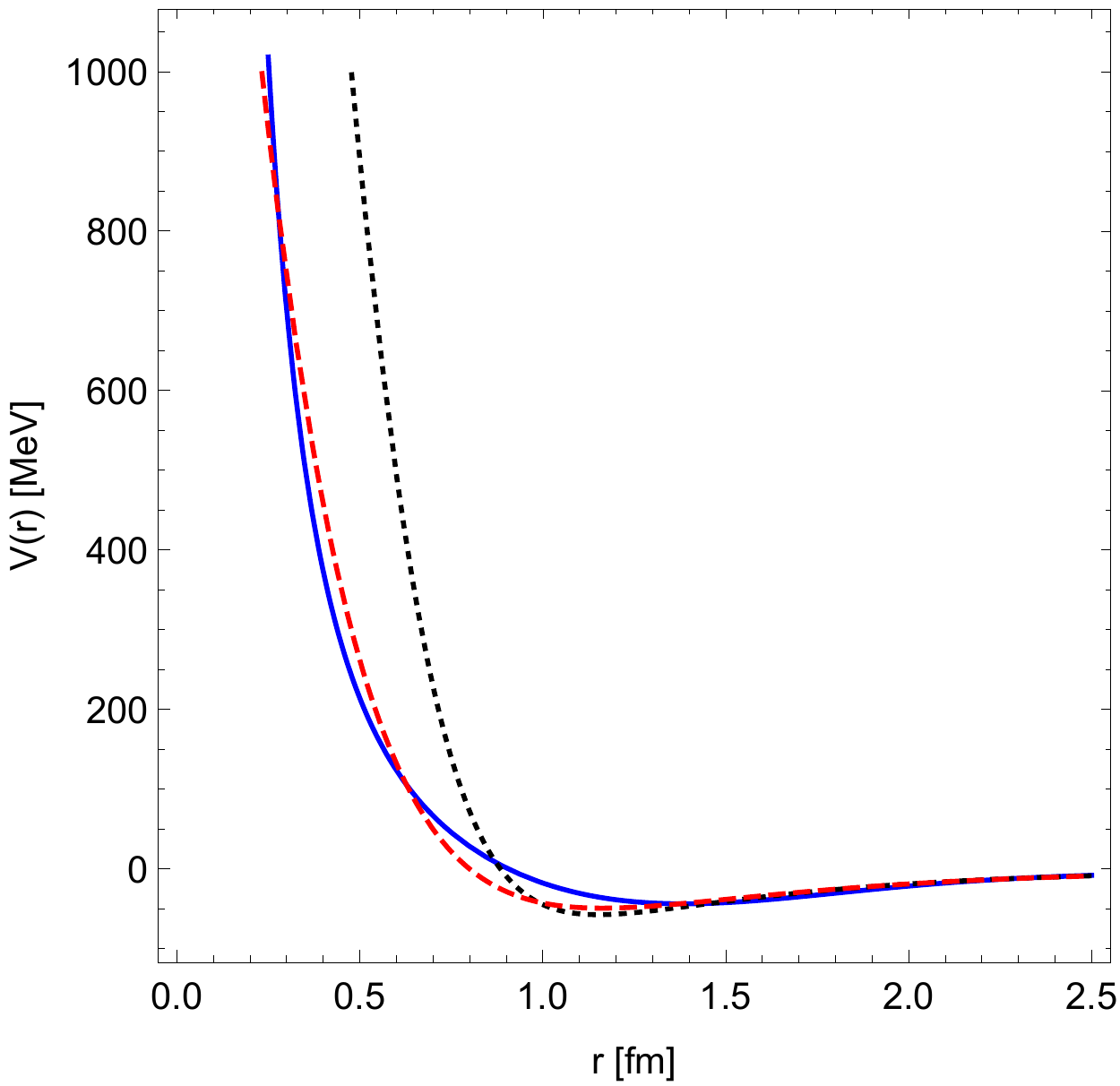}
                \caption{$\nu=2.3$}
        \end{subfigure}
        \caption{\footnotesize{Reconstruction of the $^3D_2$ channel nucleon potential
          with different singularity strength values. The solid
          (blue) line is the potential obtained with our extrapolation
                  method and the dashed (red) line is the Reid93 \cite{NuPo2}
                  potential. For comparison, the AV18 potential \cite{NuPo3}
                  is also shown (dotted, black).}}
\label{pot3D2}
\end{figure}

The results are shown in Figs. \ref{pot3P1}, \ref{pot3P0} and \ref{pot3D2}.
We found that for $^3P_1$ the best fit is $\nu=2.3$ (we also considered the
neighbouring values $\nu=2.3\pm0.2$). Similarly, for $^3P_0$ the best value
is $\nu=3.2$ and in the plots the neighbouring $\nu=3.2\pm0.2$ values are
also shown. Finally for $^3D_2$ we have chosen $\nu=2.3$, $\nu=2.3\pm0.1$.

As expected, there is no reason why the effective strength parameter $\nu$
should be integer. Indeed, in all our examples the best choice turns out to be
fractional. We summarize our results for the best $\nu$ and the corresponding $\beta_o$ values in Table \ref{Tab1}.

\subsection{An overall picture}
\label{coherent}

The nonrelativistic Hamiltonian describing the nucleon-nucleon interaction
is traditionally \cite{Reid68} written as a sum of three terms:
\beq
V(r)=V_{\rm C}(r)+V_{\rm T}(r)S_{12}+V_{\rm SO}(r)\vec S\cdot\vec L,
\eeq
where
\beq
S_{12}=3(\vec \sigma_1\cdot\vec r_o)(\vec\sigma_2\cdot\vec r_o)-(\vec\sigma_1
\cdot\vec\sigma_2)
\eeq
is the tensor operator with $\vec\sigma_{1,2}$ the sigma matrices in
the spin space of particles~1,~2 and $\vec r_o$ is the unit vector
along the line connecting the two particles. $\vec S\cdot \vec L$ is the
spin-orbit interaction operator.
Later more operators were considered, like in the Argonne potentials AV14,
AV18 \cite{NuPo3}, but here we use this simple form and further assume that
only the central and tensor potentials are singular and the total singularity
strength is described by the formula
\begin{equation}
\beta_o=A_{ST}-B_T S_{12}
\label{conj}
\end{equation}  
with the constant $A_{ST}$ depending on the total spin and isospin and
the constant $B_T$ depending only on isospin (since for $S=0$ the tensor
operator vanishes). The tensor potential needs to be singular, since
only the tensor operator $S_{12}$ has off-diagonal matrix elements in the
case of coupled channels and the singularity of off-diagonal matrix elements
of the potential is observed, for example, in the $^3S_1$-$^3D_1$ coupled problem.

The matrix elements of $S_{12}$ are as follows. For the $S=0$ uncoupled
channels ($J=\ell$) $S_{12}=0$. For the $S=1$ uncoupled channels ($J=\ell$)
$S_{12}=2$. For the coupled channels the formulas are generally more
complicated, for example, for the $^3S_1$-$^3D_1$ channel, in the $\ell=(0,2)$ basis
the matrix elements are
\beq
\begin{pmatrix}
  0&\sqrt{8}\\
  \sqrt{8}&-2
\end{pmatrix}.
\eeq
Finally, for the exceptional uncoupled channel $^3P_0$, $S_{12}=-4$.
We summarize these data in Table \ref{Tab2}.
\begin{table}[t]
\centering
\begin{tabular}[t]{c|c|c|c}
\hline
channel &$\beta_o $ value  & $S/T$ values & $S_{12}$ values 
\\[0.5ex]
\hline \hline
$^1S_0$ & 6.0 & 0/1 &
0 \\[0.5ex]
$^3P_1$ & 5.6 & 1/1 &
2 \\[0.5ex]
$^3P_0$ & 11.4 & 1/1 &
-4 \\[0.5ex]
$^3D_2$ & 1.6 & 1/0 &
2 \\[0.5ex]
$^3S_1$-$^3D_1$ & $1.6 \left( \begin{array}{cc} 2 & -\sqrt{2}  \\ -\sqrt{2} & 3 \end{array}\right)$ & 1/0 &
$\left( \begin{array}{cc} 0 & \sqrt{8}  \\ \sqrt{8} & -2 \end{array}\right)$ \\[0.5ex]
\hline
\end{tabular}
\caption{\footnotesize The values of parameters for various channels }
\label{Tab2}
\end{table}

From $^1S_0$ we get $A_{01}=6.0$ and from $^3P_1$, $^3P_0$ (approximately)
$A_{11}=7.5$ and $B_1=1.0$. The $^3D_2$ channel result only gives the linear
combination
\beq
A_{10}-2B_0=1.6.
\label{3D2}
\eeq
But since the $^3S_1$-$^3D_1$ coupled problem also corresponds to $ST=10$, the same
parameters also appear in the corresponding singularity strength matrices
\beq
\beta_o=A_{10}-B_0\begin{pmatrix}0&\sqrt{8}\\ \sqrt{8}&-2\end{pmatrix},
\quad
\nu(\nu+1)=A_{10}-B_0\begin{pmatrix}0&\sqrt{8}\\ \sqrt{8}&-2\end{pmatrix}
+\begin{pmatrix}0&0\\0&6\end{pmatrix}.
\label{numatrix}
\eeq
We have studied \cite{EZB2} this coupled problem using Bargmann-type S-matrices
and found that (neglecting small mixing effects) the
best choice is $\nu_1=1$ (s-channel) and $\nu_2=3$ (d-channel). Using these values
we can calculate the trace
\beq
{\rm Tr}\{\nu(\nu+1)\}=2A_{10}+2B_0+6=2+12=14,
\eeq
which, together with (\ref{3D2}), gives $A_{10}=3.2$, $B_0=0.8$.
Using these parameters in (\ref{numatrix}) we find the eigenvalues
$\nu_1=1.2$ and $\nu_2=2.9$. The limiting value of the mixing angle
can also be calculated.

To determine the parameters for $ST=10$ more accurately, we need to
apply the extrapolation method used in this paper to this coupled 
system. We intend to study this problem in a separate publication \cite{EZB4}.


\section{Conclusion}
\label{concl}

In this paper, motivated by recent theoretical progress which suggests a
$1/r^2$ type singular core in the two-nucleon interaction potential, we have
undertaken a systematic study of the singularity strength of this repulsion.
Starting from the experimental data for low and medium energies we
extrapolated these scattering phase shifts to high energies with a fixed
value of the singularity strength parameter $\nu$. Using Marchenko's method
of quantum inverse scattering we reconstructed the corresponding potential and
compared it to the Reid93 phenomenological potential up to the 500-1000 MeV
energy range, just above the validity of the experimental data used, to find
the value of $\nu$ giving the best match. We decided to use Reid93 because it
is known to have the best overall description of $np$ scattering data among
available phenomenological potentials. Had we compared our results to AV18,
an other successful phenomenological potential, the best values of $\nu$
would have been only slightly different and the overall pattern would have
remained unchanged.

In our work, we determine the singularity strength $\nu$ of the potential in a precise way by comparing with Reid93 potential up to relatively high energies. As it is highlighted in the generalized Levinson's theorem (\ref{gLt}), $\nu$ is a global parameter sensitive not only to high energy but also low energy scattering phase shifts. Therefore, a satisfactory global potential description, even at lower energies, can only be obtained with the true singularity strength, as it is verified especially in Figs \ref{pottest} and \ref{pot1S0}.

Our extrapolation method works for all positive $\nu$ values. This is in
contrast to the case of extrapolations based on Bargmann potentials,
frequently constructed by SUSY quantum mechanics methods, where $\nu$ can only
be integer. In the best studied example, the case of the $^1S_0$ channel central
potential, the singularity strength turns out to be $\nu=2$, both in the case
of SUSY extrapolation or using our method. But in all other examples we studied
here the best values for $\nu$ are actually fractional.

Comparing the results for $\nu$ for various channels, an overall coherent
picture emerges. We have formulated a conjecture (\ref{conj}), which expresses
$\nu$ in terms of a few parameters, depending on the spin and isospin quantum
numbers. We do not have a sufficient amount of data to determine and test these
parameters for all channels, but our conjecture seems consistent with the
singularity strength parameters even in the $^3S_1$-$^3D_1$ coupled channel problem,
found previously by Bargmann-type extrapolation. We intend to extend our
extrapolation method to coupled channel problems to be able to study these
questions further.


\vspace{5ex}
\begin{center}
{\large\bf Acknowledgments}
\end{center}

This work was supported by 
the National Natural Science Foundation of China (Grant No. 11575254),
by the Chinese Academy of Sciences President's International Fellowship
Initiative (Grant No. 2017PM0045 and Grant No. 2017VMA0041) and
by the Hungarian National Science Fund OTKA (under K116505). 
J.~B. would like to thank the CAS
Institute of Modern Physics, Lanzhou, where most of this work has been 
carried out, for hospitality. 

\par\bigskip



  


\vfill\eject

\end{document}